\documentclass{aastex}
\usepackage{emulateapj5}

\newcommand{\halpha}{H$\alpha$}

\shorttitle{SIRTF Nearby Galaxies Survey}
\shortauthors{Kennicutt et al.}

\newcommand{\Ha}{H$\alpha$}

\newcommand{\Ms}{M$_{\odot}$}

\newcommand{\um}{$\mu$m}
\newcommand{\m}{$\mu$m}
\newcommand{\etal}{et al.}

\newcommand{\ISO}{{\it ISO}}

\newcommand{\HI}{H\thinspace I}
\newcommand{\HII}{H\thinspace II}
\newcommand{\cbsm}{\textsc{Cubism}}

\begin{document}

\submitted{To appear in PASP, August 2003}


\title{SINGS:  The SIRTF Nearby Galaxies Survey}

\medskip
\author{Robert C.\ Kennicutt, Jr.\altaffilmark{1}, Lee Armus\altaffilmark{2},
George Bendo\altaffilmark{1}, Daniela Calzetti\altaffilmark{3},
Daniel A.\ Dale\altaffilmark{4}, Bruce T.\ Draine\altaffilmark{5},
Charles W.\ Engelbracht\altaffilmark{1},
Karl D.\ Gordon\altaffilmark{1},
Albert D.\ Grauer\altaffilmark{6},
George Helou\altaffilmark{7},
David J.\ Hollenbach\altaffilmark{8}, Thomas H.\ Jarrett\altaffilmark{9},
Lisa J.\ Kewley\altaffilmark{10}, Claus Leitherer\altaffilmark{3},
Aigen Li\altaffilmark{1},
Sangeeta Malhotra\altaffilmark{3},
Michael W.\ Regan\altaffilmark{3},
George H.\ Rieke\altaffilmark{1},
Marcia J.\ Rieke\altaffilmark{1},
H\'el\`ene Roussel\altaffilmark{9},
John-David T.\ Smith\altaffilmark{1},
Michele D.\ Thornley\altaffilmark{11}, and
Fabian Walter\altaffilmark{12}}

\altaffiltext{1}{Steward Observatory, University of Arizona, Tucson, AZ  85721}
\altaffiltext{2}{SIRTF Science Center, Caltech, Mail Code 220-6, Pasadena, CA 91125}
\altaffiltext{3}{Space Telescope Science Institute, 3700 San Martin Drive, 
Baltimore, MD  21218}
\altaffiltext{4}{Department of Physics \& Astronomy, University of Wyoming,
Laramie, WY  82071}
\altaffiltext{5}{Department of Astrophysical Sciences, Princeton University,
Princeton, NJ 08544-1001}
\altaffiltext{6}{Caltech, Mail Code 314-6, Pasadena, CA 91101}
\altaffiltext{7}{University of Arkansas, Department of Physics \& Astronomy,
Little Rock, AR 72204}
\altaffiltext{8}{NASA Ames Research Center, MS 245-3, Moffett Field, CA 94035-1000}
\altaffiltext{9}{Caltech, Mail Code 320-47, Pasadena, CA  91125}
\altaffiltext{10}{Harvard-Smithsonian Center for Astrophysics, 
60 Garden Street, Cambridge, MA 02138}
\altaffiltext{11}{Department of Physics, Bucknell University, Lewisburg, PA 17837}
\altaffiltext{12}{National Radio Astronomy Observatory, PO Box O, Socorro, NM 87801}

\begin{abstract}

The SIRTF Nearby Galaxy Survey is a comprehensive infrared imaging and
spectroscopic survey of 75 nearby galaxies.  Its primary goal is to
characterize the infrared emission of galaxies and their principal
infrared-emitting components, across a broad range of galaxy
properties and star formation environments.  SINGS will provide new
insights into the physical processes connecting star formation to the
interstellar medium properties of galaxies, and provide a vital
foundation for understanding infrared observations of the distant
universe and ultraluminous and active galaxies.  The galaxy sample and
observing strategy have been designed to maximize the scientific and
archival value of the data set for the SIRTF user community at large.
The SIRTF images and spectra will be supplemented by a comprehensive
multi-wavelength library of ancillary and complementary observations,
including radio continuum, HI, CO, submillimeter, $BVRIJHK$, \halpha,
Paschen-$\alpha$, ultraviolet, and X-ray data.  This paper describes
the main astrophysical issues to be addressed by SINGS, the galaxy
sample and the observing strategy, and the SIRTF and other ancillary
data products.

\end{abstract}

\keywords{galaxies: evolution -- galaxies: ISM -- ISM: dust, extinction --
stars: formation -- infrared: galaxies -- surveys}

\section{Introduction}

Nearly half of the bolometric luminosity of the local universe
is channeled through the mid- and far-infrared emission of galaxies;
this spectral region directly probes the youngest 
star-forming regions and their associated interstellar gas and dust.
However, existing infrared (IR) instruments have only begun to probe
the full ranges in star formation rates, interstellar medium (ISM)
properties, and dynamical environments found in external galaxies.
The SIRTF Nearby Galaxies Survey (SINGS) is designed to characterize 
the infrared emission of present-day galaxies and probe the full range of
star formation environments found locally,
including regions that until now have been inaccessible at
infrared wavelengths.  The observations include 
imaging and low-resolution spectroscopy
of 75 nearby galaxies ($d < 30$ Mpc),
and high-resolution spectroscopy of
their centers and a representative set of extranuclear 
infrared-emitting regions.
These data will be supplemented with an extensive library of
ground- and space-based data at other wavelengths.

The primary scientific objective of SINGS is to use observations
of these galaxies at infrared, visible, and ultraviolet wavelengths to 
obtain complete maps of the current star formation, that are 
free of most physical biases imposed by interstellar dust.
These will be used in turn to robustly model the extinction, heating, and
infrared emission of the dust, and carry out a 
critical comparison and integration of ultraviolet, \Ha, and infrared-based
estimates of star formation rates (SFRs) in galaxies.  The infrared spectra
will provide spectral diagnostics of the ionized, neutral, molecular, and 
dusty ISM phases of individual star-forming regions, and will probe the
interactions between young stars and their surrounding ISM over a much wider 
range of physical and dynamical environments than can be probed in the Galaxy.

In addition, SINGS will provide an archive of spatially-resolved images and
spectra of nearby galaxies that will enable a rich set of follow-up
investigations of star formation and the ISM.  The galaxy sample 
and observing strategy are designed to maximize the long-term scientific return
of the data for the SIRTF user community as a whole.  The multi-wavelength
archive, which will include SIRTF observations as well as a set of $BVRIJHK$,
\Ha, Paschen-$\alpha$, 
ultraviolet, far-infrared and submillimeter, X-ray, CO, HI, and
radio continuum data, will provide a foundation of
data and comparison samples for future SIRTF General Observer (GO) 
observing projects by other investigators.

This paper is organized as follows.  In \S\,2 we describe
the core science program for SINGS, and highlight some of the many
archival applications of the data set.  
In \S\,3 we describe the design and the
properties of the galaxy sample, and \S\,4 describes the SIRTF observations
themselves.  The ancillary data being obtained by our team and
by other groups are described in \S\,5.  As with all SIRTF Legacy projects
the pipeline data products will be available immediately, with no
proprietary period.  However, the SINGS team will also produce a set of
enhanced value data products (e.g., mosaicked images, spectral data cubes),
and software tools, and these are described in \S\,6.
For a general description of SIRTF and its instruments we refer
the reader to Gallagher, Irace, \& Werner (2002).

\section{Scientific Objectives}

Information on the large-scale distribution of star formation in
galaxies, and its coupling to the underlying properties of the ISM
is critical to a host of larger astrophysical problems--
the physical nature and origin of the Hubble sequence,
the structure and phase balance of the ISM, the interpretation
of observations of the high-redshift universe, the nature
and triggering of starbursts-- and as a lynchpin
for the physical understanding and modeling of galaxy formation and
evolution.  Nearby galaxies offer the unique opportunity to study this
SFR--ISM coupling on a spatially-resolved basis, over
large dynamic ranges in gas density and pressure, metallicity, dust
content, and other physically relevant parameters.

\begin{figure*}
\epsscale{2.0}
\plotone{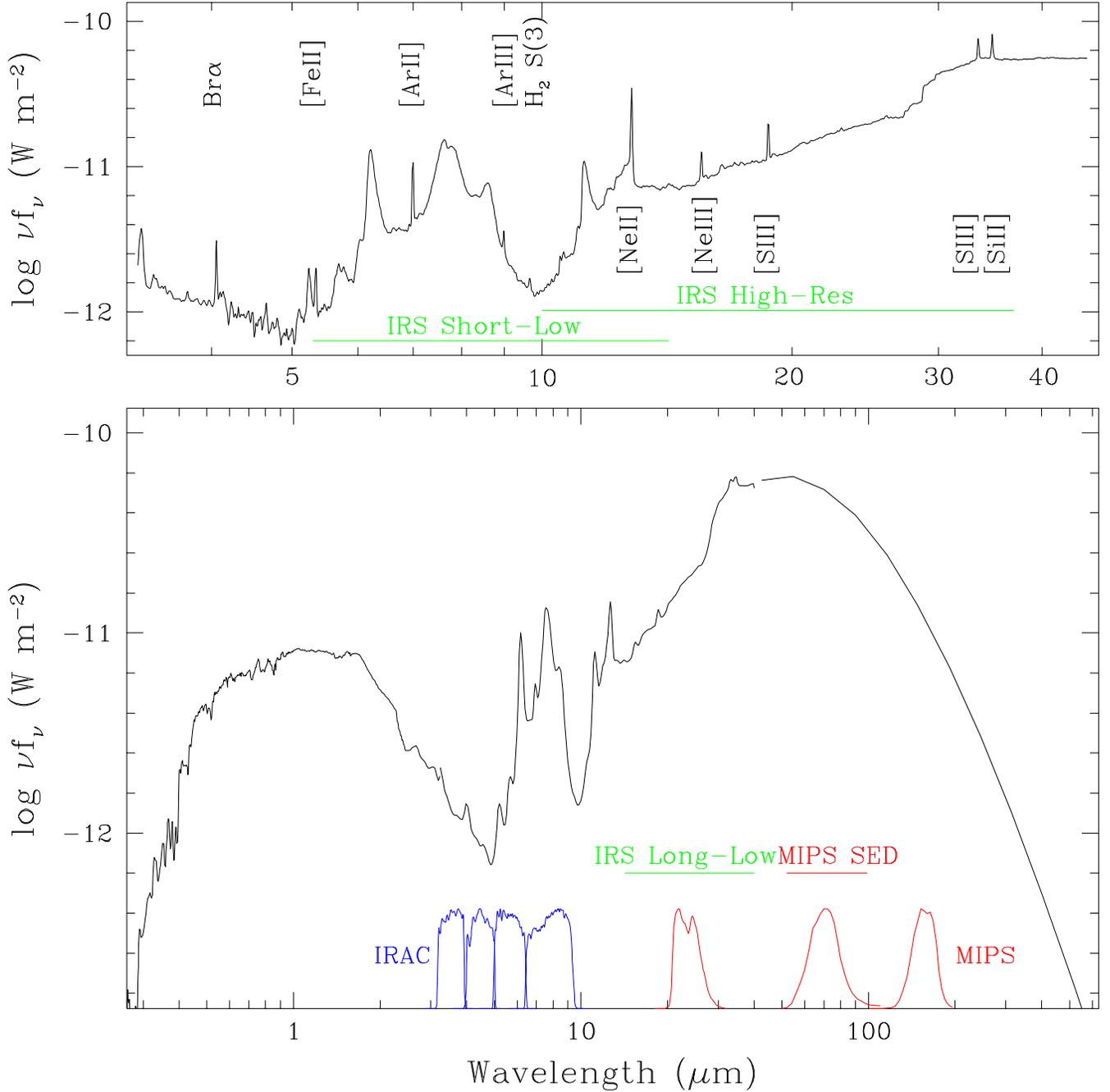}
\caption{Top:  Infrared spectrum of M82 as observed by ISO, smoothed
to the resolution of the SIRTF observations to be obtained by the
SINGS project.  Bottom:  A schematic spectral energy distribution of
the same region, extending from the near-ultraviolet to submillimeter, with
the infrared data smoothed to the low-resolution spectral mapping
mode of SIRTF.  The corresponding imaging bandpasses are also shown.}
\epsscale{1.0}
\end{figure*}

SIRTF's unique contribution derives from its ability to probe the flow of 
energy from stars into the ISM on scales ranging from the immediate
environs of star-forming molecular clouds to the radiative transfer
of diffuse radiation on kiloparsec scales (Helou 2000).  
Within star-forming regions, stellar radiation shining on the surrounding gas
produces a series of ISM components, all of which are probed in the infrared.
(Figure 1).  Ionizing radiation produces an \HII\ region that is traced by
a suite of mid-infrared fine structure lines from species covering
a wide range of ionization states.  Outside of the ionization front
the gas becomes optically thick to hydrogen-ionizing
photons, but ultraviolet photons in the range 6 eV $< h\nu <$ 13.6 eV
dissociate molecules, ionize atoms like C, Si, and Fe, and heat
the gas via grain photoelectric heating to form a photodissociation
region or PDR (Hollenbach \& Tielens 1999).
Deeper within the clouds, the warm molecular interface is
directly observable via the pure rotational transitions of molecular 
hydrogen (H$_2$).  Dust grains are present throughout these phases and provide
diagnostics of the physical conditions in the star-forming regions,
through the thermal continuum emission and the aromatic emission bands.
Much of this emission arises from
dust grains in HII regions and PDRs, but dust in ambient
clouds illuminated by the general interstellar radiation field (ISRF)
can also be important (Lonsdale \& Helou 1987).
Another energy channel -- mechanical
energy transfer -- can be traced via shock-sensitive features such as
H$_2$, [Si\,II], and [Fe\,II].

\subsection{Science Drivers:  The Physics of the Star Forming ISM and Galaxy
Evolution}

In contrast to the study of the formation of
individual stars, which has advanced dramatically over the past
two decades, our theoretical understanding of star
formation on galactic scales remains relatively immature.
The limitations begin at the very foundation, with our limited ability
to measure and map the large-scale star formation rates (SFRs) in galaxies.
Ground-based observations using \Ha\ and other nebular recombination
lines and direct observations of the ultraviolet continua of galaxies
from space have provided high-resolution maps of star formation in hundreds
of nearby galaxies (Kennicutt 1998a and references therein).  
However the interpretation of these data is limited by interstellar dust,
which absorbs roughly half of the emitted ultraviolet (UV) 
and visible radiation of 
typical star-forming galaxies, and more than 90\% in extreme cases such 
as infrared-luminous starburst galaxies 
(e.g., Calzetti et al.\ 1994; Wang \& Heckman 1996; 
Buat et al. 1999, 2002; Sullivan et al. 2000; Bell \& Kennicutt 2001,
Calzetti 2001; Goldader et al.\ 2002).  
Since the intensity of the star formation itself is strongly
correlated with the column densities of gas and dust (Kennicutt 1998b),
the most extincted regions preferentially occur in the regions with
highest SFR density.  Thus our picture of galactic star formation based on 
visible-UV observations alone is both incomplete and physically biased. 
This bias may be even more important at high redshift, where  
intense and compact starbursts associated with the
formation of galactic spheroids are prevalent 
(e.g., Barger et al.\ 2000; Giavalisco 2002).

A similar physical dichotomy between UV-bright and IR-bright regions 
is seen on the scale of individual
star-forming regions (Kennicutt 1998a,b).
The giant OB/HII associations in disks which dominate the
visible star formation have mean emitting gas
densities of order 10--100 cm$^{-3}$, neutral $+$ molecular
column densities of order $10^{20} - 10^{22}$ cm$^{-2}$
and visual extinctions of 0--3 magnitudes.  By contrast the regions
that contain the bulk of star formation in IR-luminous starburst galaxies
are located in circumnuclear environments, 
with typical densities of order $10^3 - 10^6$ cm$^{-3}$,
molecular column densities of $10^{22} - 10^{25}$ cm$^{-2}$, and
visual extinctions of 5--1000 magnitudes (Kennicutt 1998a).
Visible and UV diagnostics cannot be applied reliably 
in these regions, and instead most of our information comes from IR
and radio recombination and fine-structure lines.  Because of this
observational segregation between the physical regimes that have been 
(in)accessible
in the UV--visible and IR, the two sets of spectral diagnostics have not been
fully tested against each other, and questions remain about whether we
are interpreting the respective sets of spectra in a consistent manner.
The long-standing controversy over whether the initial mass function 
(IMF) of young stars in
IR-luminous starbursts is radically different from that observed
in optically-selected star-forming regions is a prime example of
this ambiguity (e.g. Rieke et al. 1993; Lutz 2000; 
Alonso-Herrero et al.\ 2001).
An important step toward breaking this impasse is to apply a homogeneous set
of IR spectral diagnostics across the entire range of star-forming regions,
ranging from the most dust-obscured nuclear regions to objects in the
disk that have traditionally been studied using optical spectra.

\subsection{Contributions from IRAS and ISO}

The Infrared Astrononical Satellite (IRAS) 
survey detected the integrated dust emission from tens of thousands
of nearby galaxies, and revealed a new class of infrared-luminous
starburst galaxies (Soifer, Neugebauer, \& Houck 1987).  
Since that time several groups have combined IRAS data for nearby galaxies 
with UV and/or \Ha\ fluxes, to test the reliability of the far-infrared
(FIR) luminosity
of a galaxy as a quantitative SFR tracer, and to quantitatively assess
the effects of extinction on the UV and optical SFR tracers (e.g.,
Helou 1986; Lonsdale \& Helou 1987; Rowan-Robinson \& Crawford 1989;
Devereux \& Young 1990; Sauvage \& Thuan 1992; Buat \& Xu 1996; 
Walterbos \& Greenawalt 1996; Bell \& Kennicutt 2001; 
Buat et al.\ 2002; Dopita et al.\ 2002;
Kewley et al.\ 2002).  The FIR has proven to be a uniquely powerful tracer
of the SFR in galaxies dominated
by young stars and with high dust optical depths, where the infrared continuum
provides a near-bolometric measurement of the stellar radiation
output and SFR (Kennicutt 1998b).  

Interpreting this emission in normal
galaxies has proven to be more difficult, however.  There the dust can 
also be heated by older stars, and the fractional
contribution of this ``cirrus" component to the total FIR emission
is a strong function of galaxy properties and the SFR itself
(Helou 1986; Lonsdale \& Helou 1987; Rowan-Robinson \& Crawford 1989;
Walterbos \& Greenawalt 1996).  Moreover, the infrared
emission may under-represent the output of regions with low dust content.  
Despite these concerns recent work comparing the reddening-corrected
\Ha-based and FIR-derived SFRs show a surprisingly good agreement over
a wide range of galaxy types and SFRs, which suggests that at least for
normal galaxies the effects of older stars and dust opacity on the 
FIR are less serious than anticipated
(e.g., Dopita et al.\ 2002; Kewley et al.\ 2002).
Unfortunately it is difficult to model the effects of dust extinction
and dust heating in a realistic way, because of the complex geometry
of the dust and heating sources, and the absence of high-quality
FIR data.  The best spatial resolution achievable in the far-infrared
IRAS images is of the order of an arcminute, and for all but the 
nearest galaxies this is insufficient to robustly model the dust heating
and radiation reprocessing.

Major inroads into addressing these problems have come from the 
the Infrared Space Observatory (ISO), which
provided higher-resolution imaging as well as spectroscopic observations
of galaxies in the 3--200 \micron\ range (Kessler et al.\ 1996).
Several surveys directed at nearby normal galaxies were carried out,
including the ISO Key Project on the Interstellar Medium of Normal Galaxies 
(Helou et al. 1996), which obtained comprehensive imaging and spectroscopy
for a sample of 60 galaxies with a wide range of infrared and optical
properties (Dale et al.\ 2001; Malhotra et al.\ 2001; Lu et al.\ 2003).  
Other major programs included an imaging survey of 77 
Shapley-Ames spiral and S0 galaxies (Bendo et al.\ 2002a, b, 2003), surveys
of the Virgo cluster (Boselli et al.\ 1998; Tuffs et al.\ 2002; 
Popescu \& Tuffs 2002; Popescu et al.\ 2002), and an imaging and 
spectroscopic survey of normal and barred spiral galaxies 
(Roussel et al.\ 2001a).

\begin{figure*}
\epsscale{2.0}
\plottwo{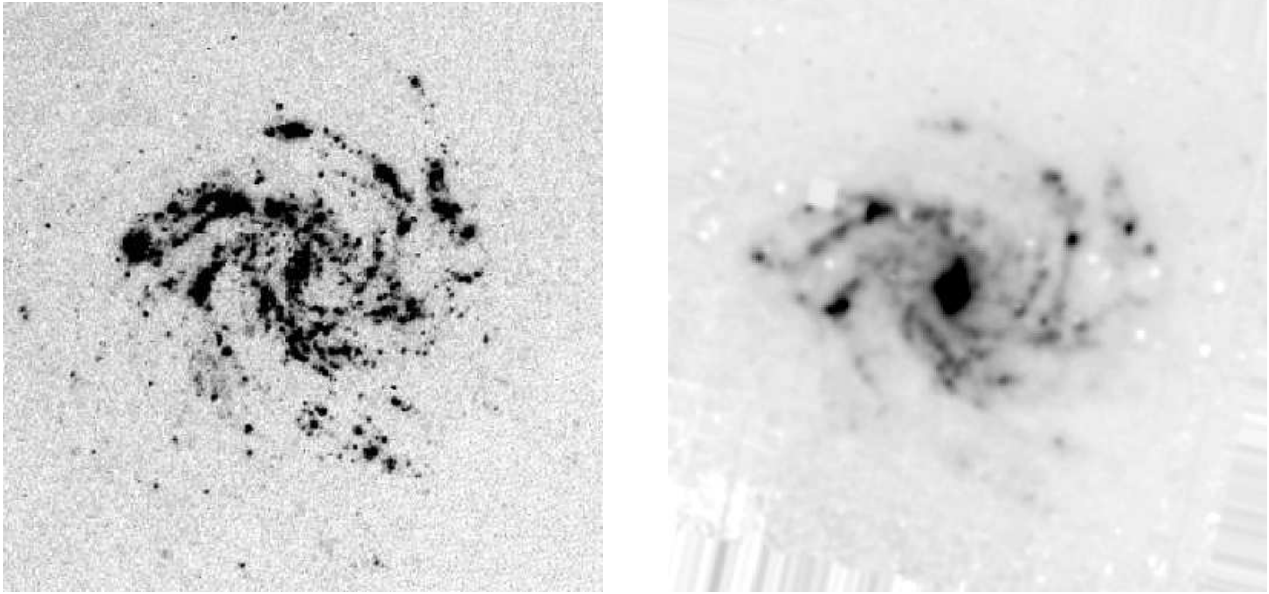}{fg2b.epsi}
\caption{The nearby (d = 5.5 Mpc) Sc galaxy NGC~6946 imaged in \halpha\ (left)
and imaged at 15 \micron\ by ISO (right).  SIRTF will produce maps
with resolutions of $<$2\arcsec at 3.5--8 \micron,
and resolutions of 6\arcsec\ and 18\arcsec\ FWHM at 24 and 70 \micron,
respectively.}
\epsscale{1.0}
\end{figure*}

The superior spatial resolution of ISO relative to IRAS provided valuable
new information on the distribution of emitting dust and the associated
star formation in galaxies, but this advantage was mainly 
limited to wavelengths shortward of 15 \um, where the dust emission
is dominated by band emission from small aromatic hydrocarbon
dust grains (e.g., Cesarsky 
et al.\ 1996; Helou et al.\ 2000; Sturm et al.\ 2000; Hunter et al.\ 2001;
Roussel et al.\ 2001a, b).  
An example is shown in Figure~2, which compares an ISO map of the
nearby galaxy NGC~6946 (a member of the SINGS sample) with a groundbased
\Ha\ image (Dale et al.\ 2000).  It is clear that in the outer disk the \Ha\
emission and dust emission trace the same general star formation component.
However there are large local variations in the infrared/\halpha\ intensity
ratio, which are due to changes in dust opacity, radiation field intensity,
and differences in the ages of the star-forming
regions.  The emission from the central starburst is much
stronger in the infrared (relative to the visible), because of the much 
higher optical depth of dust in that region.  Because the local intensity of
star formation itself tends to correlate with gas and dust column density
(e.g., Kennicutt 1998b; also see Fig.\ 4), the infrared emission provides
a probe of the most intense starbursts that is virtually inaccessible in
the visible and ultraviolet.

\begin{figure*}
\epsscale{2.0}
\plotone{fg3.epsi}
\caption{ISO SWS spectra of M82 and the Circinus galaxy, illustrating
the richness of the emission-line and molecular band spectra in this
spectral region.  Credit: ESA/ISO, SWS, Moorwood et al.\ (1996),
F\"orster Schreiber et al.\ (2003).}
\epsscale{1.0}
\end{figure*}

ISO has also made a major contribution by producing 
mid- and far-infrared spectra for a sample of infrared-luminous
galaxies, and for a few lower-extinction star-forming regions in disks
(e.g. Sturm et al.\ 2000; Thornley et al.\ 2000; Helou et al.\ 2000;
Malhotra et al.\ 1997, 2001).  The richness of emission-line and
emission-band diagnostics in this
region of the spectrum is illustrated in Figure~3, which shows 
ISO SWS spectra of M82 and the Circinus galaxy with the principal
diagnostic lines labeled.  These include ionized tracers of stellar
photoionized HII regions (e.g., [NeII]~12.8$\mu$m, [NeIII]~15.6$\mu$m,
[SIII]~18.7,~33.5$\mu$m, [SIV]~10.5$\mu$m), high-ionization tracers
for application to AGNs, shocks, and very hot stars (e.g., [OIV]~25.9$\mu$m,
[NeV]~14.3$\mu$m), low-ionization tracers of shocks and PDR regions
([FeII]~26.0$\mu$m, [SiII]~34.8$\mu$m), the S(0) and S(1) molecular
rotational lines of H$_2$ (17 and 28 \um),
and the PAH dust bands in the 5--14 \micron\ region.
Spectroscopy of [CII]~158~\micron\ in normal star-forming galaxies
has provided insight into the microphysics of the ISM.  Galaxies with
warmer far-infrared colors (larger 60/100~\um\ flux ratios) 
show a smooth decline in the [CII]/FIR ratio 
(Malhotra et al.\ 2001), which has been attributed to less efficient
photo-ejection of electrons from increasingly charged small grains in 
more actively star forming regions.  This interpretation is supported
by a strong correlation between [CII] line emission and 5--10~\micron\
dust emission in star-forming galaxies (Helou et al.\ 2001), because
the latter is a measure of the small grain photoelectric heating rate.

In terms of ISO continuum spectroscopy, we have learned that the
mid-IR spectrum ($\lambda<12$\micron) is nearly shape-invariant on the
scale of galaxies, as a consequence of the single-photon heating nature
of the PAH dust (Draine \& Li 2001).  The shape of the complete 
infrared spectrum can be
parametrized by a single variable, the 60/100~\um\ flux
ratio (Helou et al.\
2000; Dale \& Helou 2002; Lu et al. 2003).  Both of
these discoveries bode well for interpreting high-z galaxies using IR
spectral templates based on nearby galaxies.

\subsection{SINGS Project Goals}

SIRTF promises a breakthrough in this field, because it will produce 
spatially-resolved infrared spectral energy distributions (SEDs)
of galaxies with sufficient angular and 
spectral resolution to separate individual dust emission and heating 
components.  The SIRTF images have sufficient sensitivity to map the
thermal dust emission over most or all of the optical extent of spiral
and irregular galaxies.  Likewise the spectroscopic sensitivity of SIRTF
makes it possible to apply a consistent set of diagnostics
across the local populations of galaxies and star-forming regions.

To contribute optimally to these investigations, SIRTF must be used in a
manner that is best adapted to its finite angular resolution.  We have
therefore adopted a two-part strategy.  In the first part, 75 galaxies
will be imaged completely with IRAC and MIPS, and partially mapped with
low-resolution spectroscopy, to characterize the global infrared and
star-forming properties of nearby galaxies, and explore their
dependencies on type and other integrated properties. In the second
part, we will use complete spectroscopic observations
to physically characterize a set of
discrete infrared-emitting
sources.  These include the nuclear regions of
each of the 75 galaxies, and observations of 75 extranuclear regions
in $\sim$20 of the galaxies, divided between visible star-forming
regions and infrared-bright objects discovered with our SIRTF images,
chosen to contain the complete range of physical characteristics.  The
combination of these two data sets will fully characterize
the infrared properties of normal galaxies, and enable a broad set of
astrophysical applications.

A major focus of our analysis will be to test and develop
improved diagnostics of SFRs in galaxies, based on a critical comparison
of the ultraviolet, \Ha, and infrared data (e.g., Popescu et al.\
2000; Roussel et al.\ 2001a; Bell et al.\ 
2002).  The key to breaking the degeneracies that currently limit
the modeling of extinction and dust heating is to map the re-radiated
dust emission on angular scales that are comparable to those of 
individual infrared-emitting (and star-forming) regions.  
In most star-forming galaxies the bulk of the dust radiation
is emitted in the far-infrared, with a peak between 50 and 150 \micron\
(Figure~1 and Dale et al.\ 2000; Dale \& Helou 2002; Bendo et al.\ 2002b).
Hence a central element of our program is imaging of the
SINGS galaxies at 24--160 \micron\ with MIPS, and SED mapping at
10--99 \micron\ with IRS and MIPS.  At the shorter wavelengths these
maps will possess comparable or superior resolution to the ISOCAM
15 \micron\ maps illustrated in Figure 2, and even at the longest
wavelengths we will obtain images with over 200 resolution elements for  
a median-size galaxy in our sample ($D_{25} \sim$ 10\arcmin,
where $D_{25}$ denotes the isophotal diameter from the RC3:
de Vaucouleurs et al.\ 1991).
The IRAC camera on SIRTF will allow us to measure the mid-infrared
dust emission at $\sim$2\arcsec\ resolution.  Although this spectral
region (5--8 \micron) is roughly a factor of ten shortward of
the thermal dust emission peak, and the radiation is dominated by
band emission from molecular aggregates or small grains, studies with 
ISO reveal a surprisingly good correlation between the dust emissivity 
at these wavelengths and the SFR, over an extended range of physical
conditions and spatial scales (Roussel et al.\ 2001a).  We will 
test the reliability of the mid-infrared
emission as a quantitative SFR tracer, as it offers the tantalizing 
prospect of mapping star formation in galaxies on arcsecond angular
scales.

The products of this analysis will be maps of the total star formation,
as well as maps of the UV--visible extinction and separate distributions
for the UV-bright and dust-obscured star formation.  These will allow
us to delineate the broad trends in SFR as functions of local and
global galaxy properties, and they will reveal any systematic differences
between the distribution and amounts of obscured {\it vs.} 
unobscured star formation that may be present, as functions of 
radial location, metallicity, galaxy type, or dynamical environment
(e.g., Kennicutt 1983, 1998a; Roussel et al.\ 2001b).  
Since the infrared-bright
regions preferentially trace the densest and youngest star-forming 
environments, such patterns must be present.  

\begin{figure*}
\epsscale{1.75}
\plotone{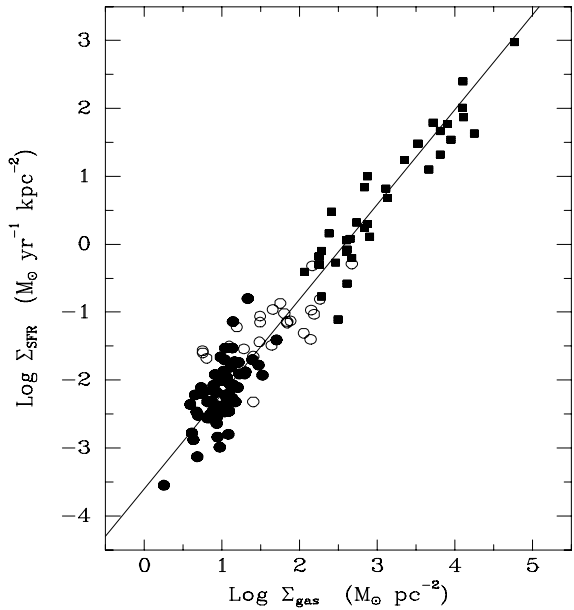}
\figcaption{Relationship between the disk-averaged star formation rate per
unit area and disk-averaged total gas density.  Solid circles are normal
star-forming disk galaxies, with SFRs derived from extinction-corrected
\halpha\ measurements, and the open circles show the corresponding values
for the centers of these galaxies.  The solid squares are infrared-selected 
circumnuclear starbursts, with SFRs determined from far-infrared emission. 
The solid line is a best-fitting Schmidt law with index $N = 1.4$.  Figure
taken from Kennicutt (1998b), and reproduced with permission.}
\epsscale{1.0}
\end{figure*}

Another area of investigation will be the combination of the
SFR distributions with spatially-resolved CO and HI maps to probe
the form of the star formation law over the full range of ISM environments
found in nearby galaxies.  The potential of such an analysis is illustrated
in Figure 4 (taken from Kennicutt 1998b), which compares the disk-averaged
SFRs and gas densities in normal galaxies (symbol) and infrared-selected
starburst galaxies (symbol).  There appears to be a single Schmidt law
(index N $\sim$ 1.4 -- 1.5) which reproduces the SFRs across this range
of regimes.  We will combine the spatially-resolved SFRs maps with
gas density data from the Berkeley Illinois Millimeter Array Survey
of Nearby Galaxies (BIMA SONG) project (Helfer et al.\ 2003) and 
high-resolution HI maps (see \S~5.5) to test
whether this law holds on a point-by-point basis.  Any such application
is limited by the validity of the CO emissivity as a molecular gas
tracer, of course, and the dust masses derived from the SIRTF SED maps
should enable us to place some useful limits on the variability of the
CO/H$_2$ conversion factor.
Other applications will include direct mapping of the dust density
and temperature distributions (e.g., Dale et al.\ 2000; Li \& Draine 2002),
construction of an improved radiative transfer model for dust in
galaxies using the DIRTY code (Gordon et al.\ 2001; Misselt et al.\ 2001), 
and use of the
unprecendented sensitivity of SIRTF to search for dust in gas-poor
galaxies, metal-poor objects, and in the halos of galaxies.  

SIRTF will also allow us for the first time to apply the
powerful suite of mid-infrared spectral diagnostics to galactic IR-emitting
components that span the full range
of metallicities, densities, extinctions,
and radiation field properties found in normal galaxies.  These
observations provide the link between the
observed infrared spectra of galaxies and the physical conditions in
the emitting gas and dust, and are crucial inputs for interpreting
the integrated SEDs and spectra of high-redshift objects
and more luminous starburst galaxies.  They will also allow us to 
study the dependence of the physical and chemical properties of the
dust (e.g., size distribution, composition) on environmental conditions 
such as metal abundance and SFR.


As a quantitative illustration of the sensitivity of SIRTF for these
applications, we can apply the nominal pre-launch instrumental 
sensitivity estimates for a few applications of interest.
For a galaxy at a fiducial distance
of 3.5\,Mpc (the distance of the M81 group, a primary component of our
sample), the spatial resolution of the telescope projects to linear dimensions
ranging from 40\,pc (2 pixels, 2.5--8\,\um) to 100, 300, and 700\,pc
($\lambda$/D respectively at 24, 70, and 160\,\um).
At this distance a 10\,s set of MIPS integrations at 24, 70,
and 160\,\um\ will detect with 10-$\sigma$ significance a star
forming region with SFR = $2 \times 10^{-4} M_\odot$~yr$^{-1}$ (comparable
to the Orion core), with a typical cloud mass (gas $+$ dust)
of order $10^5$\,\Ms.  Likewise
a 10\,s integration set with IRAC can detect the dust emission from
$\sim$10$^4$~\Ms\ of ISM illuminated by the average ISRF.
These cloud masses correspond to limiting gas column densities over
the respective resolution elements of order a few times $10^{19}$
to $10^{21}$ H~cm$^{-2}$, with deeper limits achievable with longer
integration times.
For high-resolution spectroscopy, IRS would measure the [NeII]~12.8~\um\
emission line from a Orion-luminosity region at 10-$\sigma$ significance
in a 30\,s integration, or the H$_2$ emission
from the S(0) and S(1) lines at 17 and 28~\um\ of a $5 \times 10^5$
\Ms\ cloud at 100~K in 500\,s (again at 10$\sigma$ significance).

These numbers serve to illustrate the wide range of applications that
will be possible either with the SINGS archive or follow-up GO observations.
When the IRS spectra are combined with the IR SEDs and UV-visible spectra 
we will be
able to carry out powerful cross-checks on the SFR and dust properties
of the regions as diagnosed by the independent tracers.
Specific investigations will include the use of the 
aromatic features in emission (AFE's) to constrain dust heating and
radiative transfer models (e.g., Boselli et al.\ 1998; Helou et al.\ 2000,
Li \& Draine 2001, 2002),
and as possible direct indicators of the SFR, as described earlier.
Spectra of the ionized regions will be used to constrain systematic
variations in the stellar temperatures and IMFs in the galaxies,
and to test the strong mid-infrared lines as quantitative SFR tracers
(e.g., Roche et al.\ 1991).  Observations of the low-ionization 
lines (e.g., [SiII]~34.8\um, [FeII]~26.0\um) and H$_2$ rotational lines
will be used to test for systematic differences in the physical conditions
in the environments of circumnuclear and disk star-forming regions, and
to test shock, \HII\ region, and photodissociation diagostics based on
these lines, over a much larger range of physical conditions than can be 
probed in the Galaxy.  

\subsection{Other Scientific Applications and Archival Science}

As with all of the SIRTF Legacy science projects, the target samples 
and observing strategy were designed to support a wide range of scientific
applications beyond the core program described in the previous section.
A comprehensive description of the types of archival applications
of the SINGS data is beyond the scope of this paper, but a few examples
will serve to illustrate the variety of potential applications.

One of our primary data
products will be a complete galaxy SED library, including
broadband integrated SEDs covering 0.4--160\,\um, and eventually with 
extended coverage over 0.15 -- 850\,\micron\ for many objects, when our SEDs 
are integrated with observations from GALEX and JCMT (see \S5).
A spectral library of the 150 discrete sources (circumnuclear regions,
extranuclear HII regions, extranuclear IR sources) will also be produced,
covering the 5--37\,\micron\ range, and with matching optical spectra
or emission-line fluxes over the 0.36--0.7\,\micron\ range for most regions.
These data should be especially useful for interpreting data from other 
Legacy programs (e.g., GOODS, SWIRE), for
interpreting submillimeter surveys of distant galaxies,
and for planning future facilities such as
ALMA, Herschel, and SAFIR.  The same data should be 
useful for modeling the composite emission from ultraluminous
infrared galaxies (ULIRGs), and by helping to identifying the truly
unique physical attributes of these objects (not represented in
the normal galaxies in our sample), and thus constrain their 
origins and evolutionary paths.

The IRAC images at 3.6 and 4.5\,\um, when combined with groundbased
near-infrared images, and correcting for hot dust emission at these
wavelengths, can provide constraints on the stellar mass distributions
in galaxies, and these are relevant to a wide range of problems including bar
structures, spiral density waves, and better constraining the dark matter
properties of galaxies, e.g., by modeling the stellar mass--to--light
ratio.  The wide range of inclinations in the sample can also be used
to study the extinction properties of disks as a function of wavelengths,
and address the long-standing debate over the opacities of galactic disks
at visible and near-infrared wavelengths 
(e.g., Valentijn 1994; Giovanelli et al.\ 1994; Burstein 1994).

The SINGS data will also provide several means of studying the
spatial distribution of hot and cold dust in galaxies (e.g., Lu
et al.\ 2003).  The IRAC
bandpasses have been selected to allow separation of aromatic emission
band features from (largely) continuum regions, and thus it will
be possible to construct high-resolution emission maps of the warm dust.
At longer wavelengths SIRTF will offer unprecendented sensitivity to
emission from the cold ``cirrus" component of dust, and these can
be used to test for the presence of dust halos, streamers, or clouds
around nearby galaxies
(e.g., Nelson et al.\ 1998, Alton et al.\ 1998, Trewhella et al.\ 2000,
Stickel et al.\ 2003).  The MIPS images
can also be used to place strong limits on the total ISM (gas + dust)
contents of early-type galaxies.

Although it is not a main component of the SINGS core science program,
the connection between AGN activity and circumnuclear star formation 
activity can be explored in detail with these data.  The 
sample contains the full range of nuclear types, including HII region,
LINER, and Seyfert nuclei.  The availability of high-resolution spectra
in the 10--37\,\micron\ region (and low resolution at 5--14\,\micron)
will make it possible to compare the physical conditions and ionization
properties of the regions from fine-structure line diagnostics with
those derived from optical spectra.  The centers of the SINGS galaxies will be 
spectroscopically mapped at 15 positions (\S\,4), providing some separation
of nuclear and circumnuclear emission, within the limitations imposed
by the spatial resolution of SIRTF ($\sim$2--9\arcsec\ FWHM at 
5--37\,\micron). 

As a final example, a wide range of gasdynamical processes, including
large-scale shocks
produced by bars, spiral density waves, and interactions are manifested
in the mid-IR continuum and the shock sensitive [SiII], [FeII], and
H$_2$ features.
The IRAC imaging and low-resolution IRS SED strips should isolate examples
of these shocks, which can be followed up with pointed spectroscopy 
as part of the SIRTF GO Program.

\section{Sample Definition and Properties}

In order to maximize the archival value of the SINGS dataset, and
to incorporate as diverse a population of galaxies and infrared
emitting regions as possible, 
we adopted a physically-based approach to defining the samples.
These address the needs of the scientific program described
in the previous section, and also provide representative
samples for other astrophysical applications, as well as 
reference and control samples for future GO observing programs.

\subsection{Galaxy Sample}

In order to achieve the scientific goals described above and at the
same time to maximize the archival value of survey, we decided from
the outset to observe as diverse a set of local {\it normal} galaxies
as possible.  The size and precise nature of the sample, however,
was heavily influenced by a set of technical considerations.
The principal limiting factor
in the performance of SIRTF for this type of science is its angular
resolution, which ranges from about 2\arcsec\ with IRAC to 40\arcsec\
in the 160\,\micron\ imaging channel of MIPS.  To maximize
the linear resolution we need to study nearby objects.  We 
found that most of the range of normal galaxy properties could be covered
by a sample with a maximum distance of 30 Mpc; the median distance of
the sample is 9.5\,Mpc (for H$_0$ = 70\,km\,s$^{-1}$\,Mpc$^{-1}$).  
Most of the IRAC and MIPS 
imaging channels cover a field of 5\arcmin\ square, so to make efficient
use of the observing time we selected most of our galaxies to have
sizes in the 5--15\arcmin\ range 
(a few smaller galaxies were included in order to cover the full physical
parameter space described below).  Finally, the sample size
was dictated by a balance between achieving a representative range of
galaxy properties and practical limitations in observing time.
This led to a sample of 75 galaxies, requiring 512 hours of SIRTF time.

\begin{figure*}
\epsscale{2.0}
\plottwo{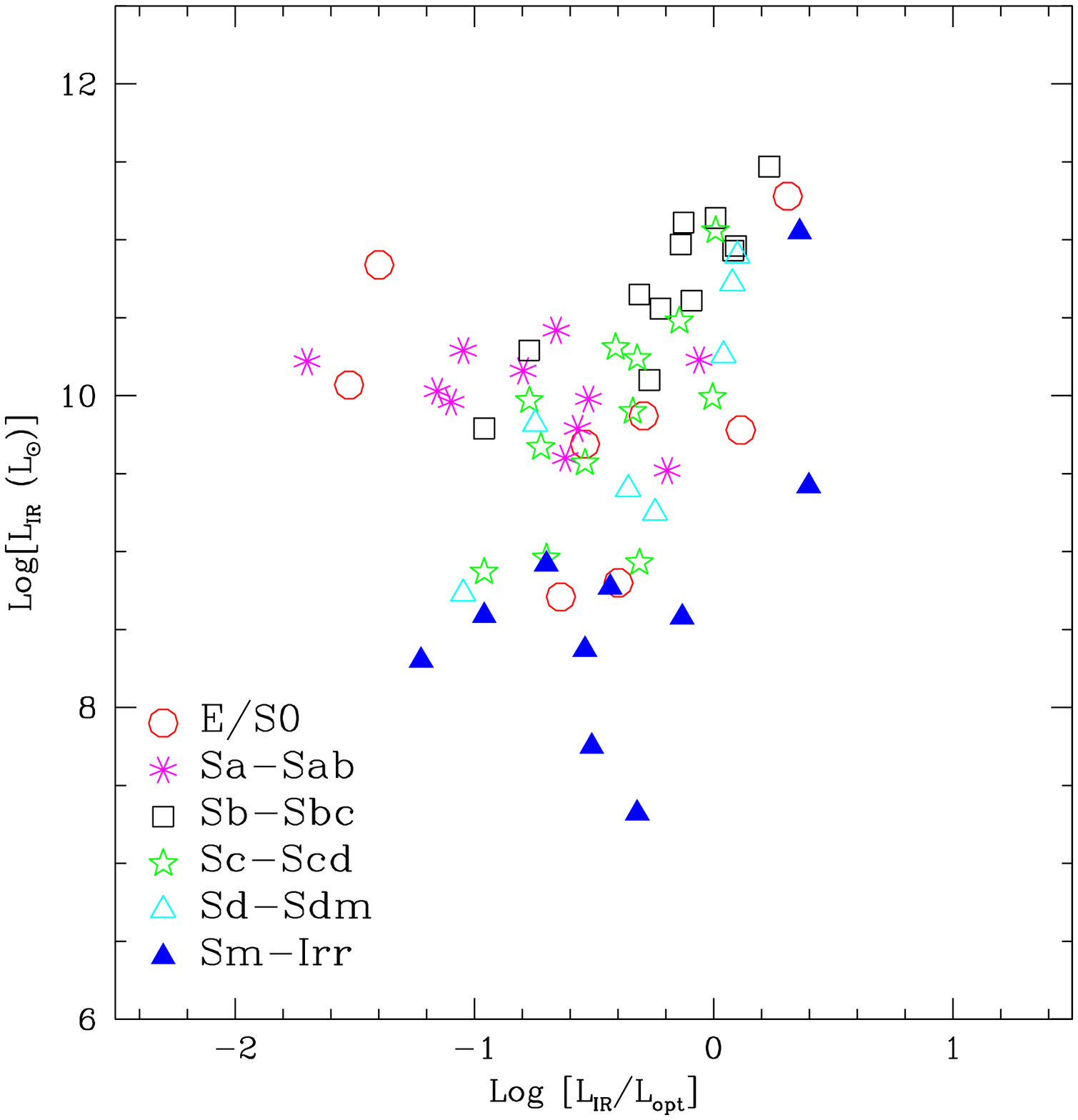}{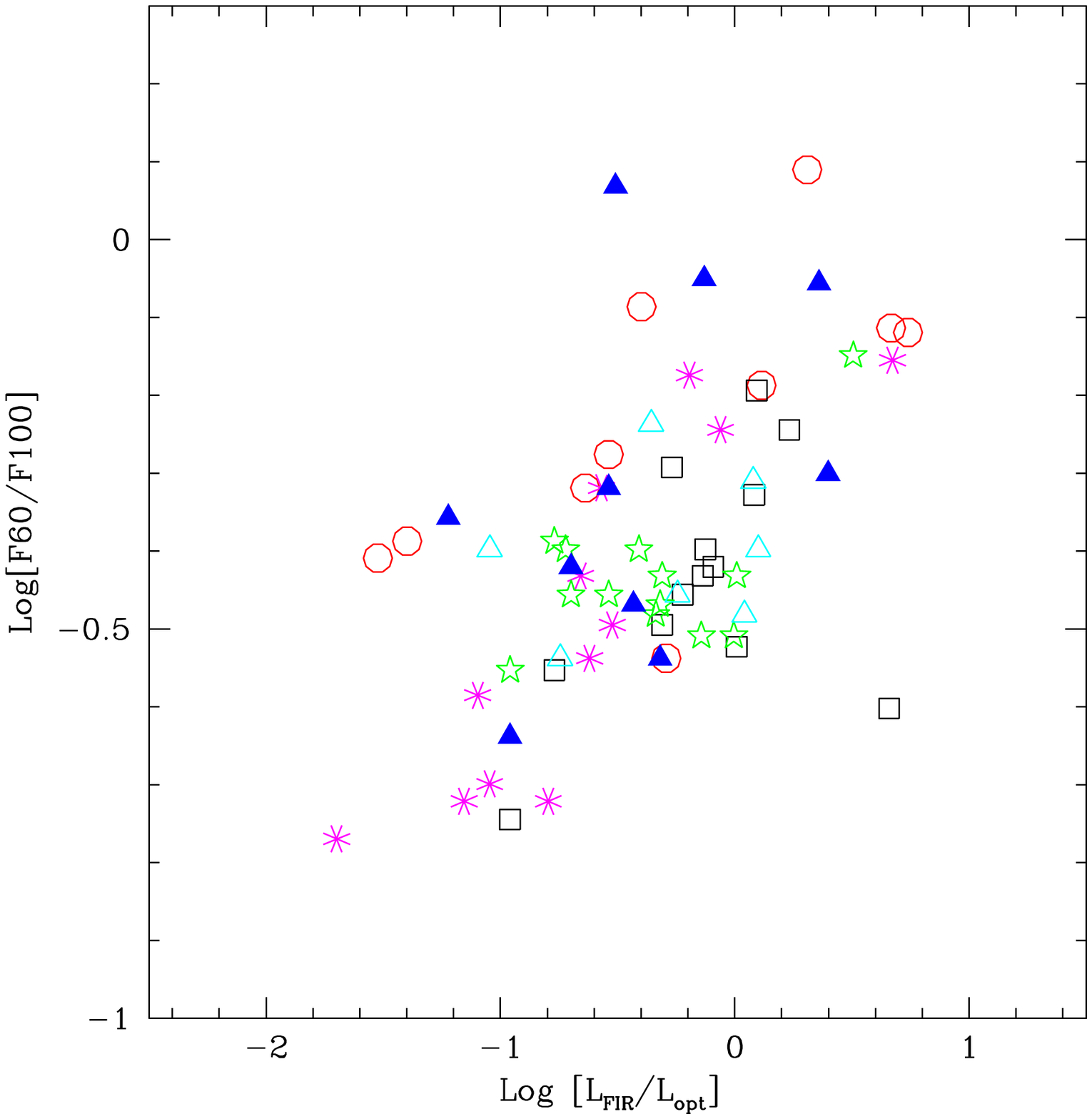}
\caption{Diagrams showing the range of galaxy properties in the SINGS
sample, in terms of morphological type (symbols), infrared luminosities,
infrared/optical ratios, and far-infrared colors.}
\epsscale{1.0}
\end{figure*}

To construct a sample of candidate galaxies we adopted a methodology
used in the ISO Key Project (Helou et al.\ 1996), and considered a 
3-dimensional parameter space of physical
properties: morphological type (which also correlates with bulge/disk 
structure, gas fraction, and SFR per unit mass), luminosity (which
also correlates with galaxy mass, internal velocity, and mean metallicity), and
FIR/optical luminosity ratio (which also correlates with the dust
optical depth, dust temperature, and inclination).
We examined the distribution of nearby galaxies in this 3-dimensional space
and selected objects that cover the full range of properties that ought
to be detectable with SIRTF.
Specifically, we chose roughly a dozen galaxies in each RC3
(de Vaucouleurs et al.\ 1991) type (E--S0, Sa--Sab, Sb--Sbc, Sc--Scd,
Sd--Sm, Im--I0), which covered the full combination of luminosity
and infrared/optical ratio.
Figure 5 shows the distribution of SINGS galaxies in this parameter space.
The sample is very diverse,
covering a factor of $10^5$ in infrared luminosity and $10^3$ in
L$_{IR}$/L$_{opt}$.  

When constructing the sample we also took pains 
to cover a representative range of other galaxy properties, including
nuclear activity (quiescent, starburst, LINER, Seyfert), inclination,
surface brightness, CO/HI ratio, 
bar structure, spiral arm structure (grand-design vs flocculent), and
environment (i.e., isolated galaxies, interacting galaxies, group
members, and cluster members).  Our goal was not to cover a representative
range of all of these secondary properties (impossible with a sample
of only 75 galaxies), but rather to avoid introducing an unintended 
bias into the sample with respect to any of these properties.

The final selection of galaxies also took into account practical factors such
as Galactic and ecliptic latitude. We excluded most galaxies near the Galactic 
plane to avoid high background at long wavelengths (from interstellar cirrus
emission), and to avoid a high density of foreground stars at the
shorter wavelengths.  We also avoided concentrating too much of the program
near the ecliptic equator, where the zodiacal background is high
and scheduling of observations is more constrained.  However 
some important regions lie near the ecliptic equator 
(e.g., Virgo cluster, Leo groups), so some observations in this region
were unavoidable.

Table~1 summarizes the range of properties in the SINGS sample,
and Table~2 gives a complete listing of the galaxies and some 
of their relevant properties.
Although our sample comprises representative cross section of galaxies  
found in the Local Supercluster, it does not include the absolute extremes 
in properties that can be found over larger local volumes.  For  
example, we did not include any examples of ultraluminous infrared
galaxies (ULIGs), luminous AGNs, cD galaxies, dwarf spheroidal galaxies, 
the most extreme metal-poor dwarf galaxies, or luminous low surface 
brightness galaxies.  This was either because the objects would not
be spatially resolved at the longest wavelengths measured by SIRTF
(e.g., ULIGs), or because ample samples of such galaxies already are 
included as reserved observations in the Guaranteed Time Observers'
(GTO) Program.  In order to maximize the science return on the SIRTF
observing time we chose to avoid duplication, in favor of eventually 
integrating any needed GTO data into future analyses.

\begin{figure*}
\plotone{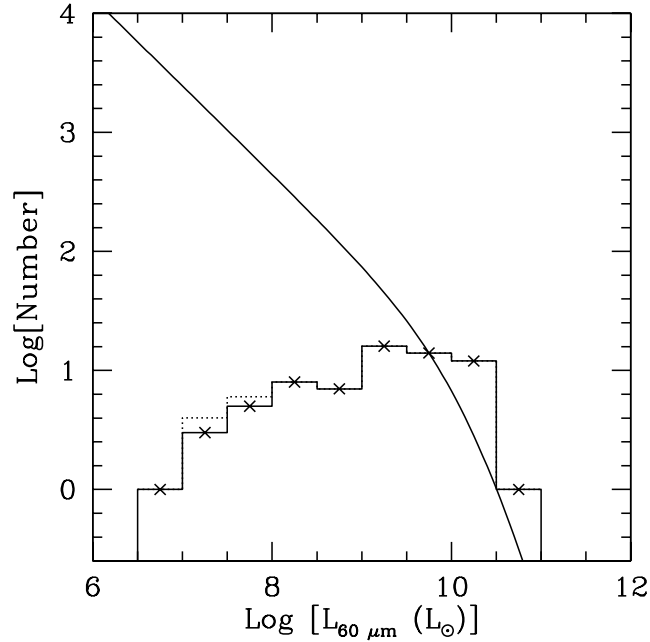}
\caption{Distribution of 60\,\micron\ luminosities for the SINGS sample.
The luminosities are calculated from the corresponding band 
IRAS fluxes (as reported in NED) and the distances listed in Table~2.
IRAS data exist for 67 of our galaxies, and upper limits for 2 (shown 
as dashed lines in the histogram).  For 6 dwarfs, there are no IRAS 
data available. The continuous line is the far-infrared luminosity 
function of Strauss \& Huchra (1988), normalized arbitrarily to match the 
high-end of our luminosity histogram.}
\end{figure*}

The deliberate emphasis on constructing a physically diverse sample also
means that SINGS should not be regarded as a 
representative, randomly drawn sample of the local universe, in 
either a volume-limited or flux-limited sense.  This point in illustrated
in Figure~6, which compares the distribution of 60\,\micron\ luminosities
of the SINGS galaxies with the local infrared luminosity function from
Strauss \& Huchra (1988).  We deliberately constructed the SINGS sample
to sample roughly equal logarithmic bins in luminosity, and this is
reflected in the flat distribution in Figure~6.  As a result the number of 
IR-faint galaxies (mostly dwarfs) is underrepresented relative to their actual
space densities.  Likewise the sample lacks any ultraluminous objects,
as mentioned earlier.  This selection should be borne in mind when
interpreting results from the SINGS sample as a whole, and particularly
when using the survey as a control sample for follow-up GO studies.
Nevertheless the sample covers the range of galaxy
types and star formation environments that are responsible for $>$95\% of
the total infrared emission in the local universe (Sanders \& Mirabel 1996).

Although it was impractical to construct a fully volume-representative
sample for SINGS, we decided to incorporate one complete luminosity-limited
subsample into the survey, namely the M81 group (Tully 1987; 
Karachentsev et al.\ 2002).  Our sample includes including M81, M82, NGC~2403, 
NGC~2976, NGC~4236, IC~2574, Ho~I, Ho~II, Ho~IX, DDO~53, DDO~165, M81 DwA, and
M81 DwB.  Four other group members, NGC~2366, NGC~3077, UGC~4486, and
VII Zw 403 are already in the reserved observation catalog, and those
data eventually will be available to complete coverage of the group
\footnote{This sample does not include a small number of recently discovered
extreme dwarf and tidal dwarf galaxies (Karachentsev et al.\ 2002; 
Makarova et al.\ 2002)}.  
The M81 group offers the advantages of proximity (d $\simeq$ 3.5 Mpc),
high visibility for SIRTF, and a diverse galaxy population, ranging
from extreme dwarfs ($M_B = -12.5$ for M81~DwB) to a large spiral
(M81) and the nearest example of an infrared-luminous starburst
galaxy (M82).  

\subsection{Spectroscopic Sample}

The spectroscopic sample is comprised of the nuclear regions of
74 of the 75 SINGS galaxies (the center of M82 is a reserved target,
to be mapped by the IRS instrument team), and 75 extra-nuclear 
regions.\footnote{Previously reserved observations of the centers of
several other SINGS galaxies were released by J.R.~Houck and M.W.~Werner
to allow them to be incorporated into our project.  We are indebted
to these individuals and their teams for their cooperation and commitment 
to the overall scientific return from SIRTF.}
The extranuclear targets are divided between $\sim$40 optically-selected
OB/HII regions and $\sim$35 infrared-selected regions, most of which
will be chosen from the SINGS images themselves.

A similar physically-based approach has been used to define the extra-nuclear
target sample.  The optically-selected \HII\ regions are located in nine
of the nearest SINGS galaxies, and were chosen to span wide ranges
in metal abundance (0.1 -- 3 $Z_\odot$), extinction-corrected 
ionizing luminosity (Q$_H = 10^{49} - 10^{52}$ photons s$^{-1}$),
visual extinction (0--4 mag), radiation field intensity (100-fold range),
ionizing stellar temperature ($T_{eff}$ = 35 -- 55 kK), and local 
H$_2$/\HI\ ratio ($<$0.1 to $>$10), as inferred from CO and 21~cm maps.
We have restricted the optically-selected \HII\ regions to those with 
extinction-corrected 
$f_{H\alpha} \ge 2 \times 10^{-16}$ W m$^{-2}$ (corresponding to 
$f_{Br\alpha} \ge 6 \times 10^{-18}$ W m$^{-2}$),
to ensure adequate detection of 
the main fine-structure lines in a reasonable exposure time.
Table 3 lists the current candidate spectroscopic targets along 
with their relevant physical properties.  

Optically-selected \HII\ regions cover only part of the full
population of bright star-forming regions and infrared sources, so
it is important to independently select a subsample of extranuclear
targets from infrared maps.  We have used published ISOCAM mid-infrared
maps to select some of the target regions listed in Table 3, and
will select at least 30 additional targets from the SINGS images,
on the basis of infrared luminosity, mid-infrared colors, and 
infrared-optical flux ratios.  This will ensure a fully representative
sampling of the SINGS extranuclear source population as a whole, 
including totally optically-obscurred objects that would never have been
selected at other wavelengths.  In addition a number of infrared-bright
circumnuclear regions will be included in the SINGS observations of
the centers of these galaxies.

\section{SIRTF Observations}

\begin{figure*}
\epsscale{2.0}
\plotone{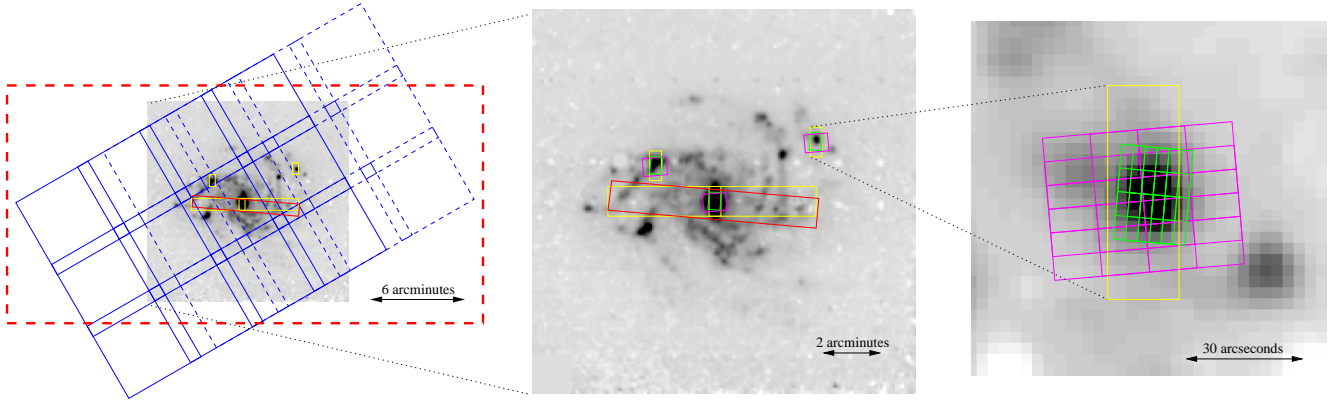}
\caption{ISOCAM image of NGC~6946 
shown at three different
scales to illustrate our observing strategy.  Left: Blue lines show the
3$\times$4 grid of IRAC imaging, and the dashed red lines show the region
to be mapped by MIPS imaging (15\arcmin$\times$30\arcmin).  Middle:
Regions to be scanned at low spectral resolution with MIPS SED mode
(red), IRS low-resolution 14--40~\um\ (large yellow box) and IRS low-resolution
5--14~\um\ (small yellow box).  Right: Expanded image of an extranuclear
region, with areas covered by IRS low-resolution 5--14~\um\ (yellow),
high-resolution 10--19.5~\um\ (green), and high-resolution 19.3--37~\um\
(violet) superimposed.  The orientations shown
are illustrative, and will differ depending on scheduling (the observations
are not dependent on a particular orientation).}
\epsscale{1.0}
\end{figure*}

In keeping with the archival design of the SINGS observing program,
we chose to maximize the homogeneity of the imaging and spectroscopic
dataset across the 75 galaxy sample.
Our basic observing strategy is depicted in Figure 7, which shows
an ISOCAM image of a SINGS target, NGC~6946, on three scales with
the footprints of the MIPS, IRAC,
and IRS observations overlaid.   For each galaxy full-coverage imaging 
(to $R > R_{25}$) will be obtained at all 7 IRAC and MIPS channels
(3.6, 4.5, 5.8, 8.5, 24, 70, 160\,\micron), with sufficient additional
sky coverage to allow for a robust background subtraction.  

\begin{figure*}
\epsscale{2.0}
\plotone{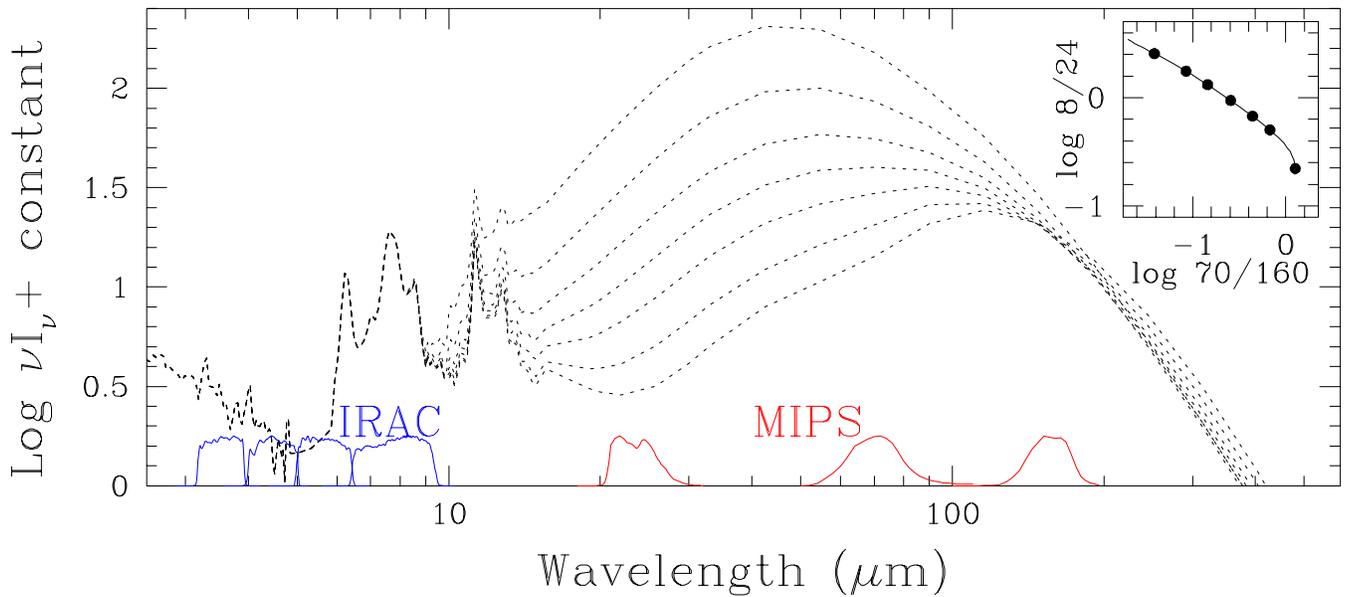}
\caption{Model SED's for normal galaxies from Dale \& Helou (2002), 
illustrating the expected range of SED shapes, and the necessity
of adding low-resolution spectra in the 14--99~\um\ region to enable the
interpretation of the IRAC and MIPS broadband colors (inset).}
\epsscale{1.0}
\end{figure*}

The coarse spectral sampling of the broadband MIPS and IRAC maps
by themselves
is insufficient to constrain the infrared SED shapes of galaxies,
much less diagnose the temperature distributions and properties of the emitting
grains, as illustrated in Figure 8.
Therefore we are supplementing our imaging with IRS 14--40\,\um\
spectral maps and MIPS 52--99\,\um\ SED mode strip scans
in all 75 galaxies.  These will permit robust modeling of the dust
continuum and heating over a several square arcminute slice of
each galaxy, and with the images help constrain the spectral shapes
throughout the disks.  Additional 
spectral maps will be obtained at 5--14\,\micron\ and 10--37\,\micron\
(at high resolution) for the centers of every galaxy except M82,
and for the 75 extra-nuclear regions described earlier. 

The remainder of this section describes the SINGS observing strategy
in detail.  This information is provided in part for the benefit of 
prospective SIRTF proposers, who may find it useful for planning their own
imaging or 
spectroscopic observations of nearby galaxies or other bright extended
sources.  A summary of the observations is given in Table~4;
readers who are only interested in the data products may wish to
skip the detailed discussion.  

\subsection{IRAC Imaging}

A majority of our galaxies are larger than the 5\arcmin\ format 
of the IRAC detectors, so we will construct
mosaics from multiple pointings to provide the necessary
field coverage.  A generous overlap (50\%) between adjacent exposures 
will provide robust cross-registration of the images, determination of
bias drifts, and cosmic ray rejection.  
After one series of exposures we will have observed all locations in 
the central part of the mosaic four times.
Because the four IRAC detectors map two non-overlapping fields of view,
mosaics of the larger galaxies will include regions above and below
which are only observed in two of the four channels.  
These regions are located well outside of the galaxies, 
and will be used to measure the sky background level.

The sensitivity of IRAC is sufficiently high to saturate the detectors
in the inner regions and in bright spiral arms, where 7\,\micron\ surface 
brightnesses of up to hundreds of MJy~sr$^{-1}$ are expected 
(Dale et al.\ 1999).
To address this problem we will observe the SINGS galaxies in a 
high dynamic range mode, in which an additional short exposure (1--2\,s)
is obtained.  This will allow us to correct for any saturated pixels 
in the longer exposures.

We will image in all four IRAC bandpasses (3.6, 4.5, 5.8, 8.0~\micron)
to at least the full optical radii
($R_{25}$) of the galaxies.  The four 256$\times$256 element 
detectors observe adjacent regions of the sky simultaneously,
each covers a region of 5\arcmin\,$\times$\,5\arcmin\
with 1\farcs2 pixels.  Galaxies with $D_{25} < 5$\arcmin\ will
be observed in a pair of four-position dithered exposure sets.  Dithered
exposures will enable us to recover somewhat more spatial information
in the reconstructed images, as described in \S\,6.1.  In order to
facilitate the detection and removal of asteroids (a significant
contamimation for targets near the ecliptic), our observations will 
be carried out in two sets of exposures separated by 1--24\,h.

A majority of our galaxies are larger than the 5\arcmin\ format 
of the IRAC detectors, so we will construct
mosaics from multiple pointings to provide the necessary
field coverage.  A generous overlap (50\%) between adjacent exposures 
will provide robust cross-registration of the images, determination of
bias drifts, and cosmic ray rejection.  
After one series of exposures we will have observed all locations in 
the central part of the mosaic four times.
Because the four IRAC detectors map two non-overlapping fields of view,
mosaics of the galaxies will include regions above and below
which are only observed in two of the four channels.  
These regions are located well outside of the galaxies, 
and will be used to measure the sky background level.
Like the smaller galaxies, each mosaiced galaxy will be observed twice
to allow for the detection of asteroids.
So there will be a total of eight exposures for all locations in the
central part of the mosaic.

In keeping with the overall design strategy of the SINGS project we
will obtain uniform exposure times (per detector field) across the galaxy 
sample, apart from small variations within individual mosaics depending
on how many times a given region of sky is covered by a particular detector.
The baseline integration time per pointing (30\,s) was chosen as a
compromise between providing acceptable signal/noise across the galaxies
and keeping the total observing time to a practical level (45\,h).
For moderate zodiacal backgrounds (the dominant source at these wavelengths)
we expect a limiting 1-$\sigma$ sensitivity of 0.04~MJy~sr$^{-1}$, which
is comparable to the expected average 7~\micron\ surface brightness of
a typical SINGS galaxy at R$_{25}$ (Dale et al.\ 2000).  
The signal-to-noise ratio improves at smaller radii and shorter wavelengths.
For example,we expect to get a signal-to-noise ratio per pixel 
of $\sim$5 for detecting stellar photospheres  at R$_{25}$.  
Since the zodiacal background varies by factors of several over 
the sky, the limiting sensitivities are expected to 
vary by about $\pm$40--50\%\ across
the sample.  However these variations will be small in comparison to
the intrinsic variations in infrared emission among the galaxies themselves.

The sensitivity of IRAC is sufficiently high to saturate the detectors
in the inner regions and in bright spiral arms, where 7\,\micron\ surface 
brightnesses of up to hundreds of MJy~sr$^{-1}$ are expected 
(Dale et al.\ 1999).
To address this problem we will observe the SINGS galaxies in a 
high dynamic range mode, in which an additional short exposure (1--2\,s)
is obtained.  This will allow us to correct for any saturated pixels 
in the longer exposures.

\subsection{MIPS Imaging}

Each galaxy in the sample will be imaged in all 3 MIPS bands (24, 
70, and 160~\micron).  
We set as a goal a minimum S/N = 3 at R$_{25}$ in all three bandpasses.
To estimate the surface brightness
levels required, we used the radial profiles of 11 nearby, face-on galaxies
observed with the IRAS satellite, and processed with the HIRES algorithm
to provide maximum spatial resolution (Auman, Fowler, \& Melnick 1990;
Kerton \& Martin 2001). 
These profiles were then used to determine the 25\,\micron\ and 60\,\micron\
surface brightness at $R_{25}$ in each galaxy.  This exercise yielded 
average values of 0.21~MJy~sr$^{-1}$ at 25\,\micron\ and 
0.51~MJy~sr$^{-1}$ at 60\,\micron, 
in very good agreement with the mid-infrared estimates above, when extrapolated
to longer wavelengths using typical galaxy SEDs (Dale et al.\ 2001).
We adopted these values as the 3$\sigma$ sensitivity goals at 24\,\micron\ and
70\,\micron. Typical galaxy surface brightnesses at 160~\micron\ are comparable
to or greater than those at 60\,\micron\ (Engargiola 1991; Bendo et al. 2002a),
so we also set 
a sensitivity goal of 0.51~MJy~sr$^{-1}$ at 160\,\micron.  
The 3$\sigma$ goals above
are equivalent to $1\sigma$ point-source sensitivities of $67~\mu$Jy, 1.9~mJy,
and 6.1~mJy, respectively.  As with the IRAC observations, variations in
zodiacal background (at 24\,\micron) and interstellar background (at 
70 and 160\,\micron) will affect the actual results, but even in the
high background regions we expect to achieve these sensitivity targets within
a factor of two.

Observations of galaxies at 70~\micron\ and 160~\micron\ impose special
challenges for observations and data processing with MIPS, because of
the extended low surface brightness structure in the sources, the
difficulty of separating physical structures from transient features
in the Ge:Ga detectors, and foreground structures such as interstellar 
cirrus in the Milky Way Galaxy.  
For that reason we describe the design of our observing and data analysis
strategies in some detail, in hopes that it may aid future SIRTF users
in designing their observing programs.

At the wavelengths covered by MIPS our observations will be confusion
limited, and the sensitivity of the detectors is sufficient to 
detect most regions in exposure times of seconds.  Consequently we have 
chosen to use short integration times (3--4~s), and multiple integrations
on each region, to 
minimize saturation effects and ensure a high level of redundancy in the data.
For expected average background levels (Young et al.\ 1998), the nominal
sensitivities given above are 
achieved in 1 photometry cycle at 70 and 160~\micron,
and is within 10\% of
being achieved after 4 cycles at 24~\micron.  Multiple
photometry cycles (2--3, depending on wavelength) will be obtained 
to ensure sufficient redundancy for removal of detector transients,
meet the sensitivity goals, enable super-resolution capability
(160~\micron), and produce consistently high-quality images.  As with
the IRAC imaging the observations will be divided into two sets 
offset by 1.5 160~\micron\ pixels, and separated 
by at least 3~h to detect asteroids.  

The MIPS instrument enables imaging to be obtained either in a conventional
``photometry" mode, or in a ``scan map" mode, in which a scanning mirror
allows for continuous integration over large areas of sky.  
We will use the small-field photometry modes for sources smaller than about
D$_{25} = 2$\arcmin, and large-field photometry modes for sources up to 
D$_{25} \sim 4$\arcmin.
Larger galaxies will be imaged using the MIPS scan map mode.
To improve sampling and mitigate the effects of bad pixels at 160~\micron\,
each scan leg will be offset from the previous one by one-half the array width,
using the $148^{\prime\prime}$ cross-scan step in both the forward and reverse
directions.  This imaging strategy ensures that each point on the sky will be
observed by at least 40 (4 at 160~\micron) independent pixels.  Due to the
Ge:Ga transient behavior, we feel that 4 observations are the minimum
required to generate high-quality data.

\subsection{MIPS SED Observations}

We will perform a small raster of SED mode observations (52--99~\micron) in a
radial strip in each galaxy, designed to overlap with the IRS low-resolution
strip scans (below).  In most galaxies (D$_{25} \le 10$\arcmin), this raster 
will
consist of 7 positions, overlapped by half the slit width to cover in full a
$1\arcmin\times3\arcmin$ region.  In galaxies larger than 10\arcmin, a second,
similar set of SED observations will be added to produce a
$\sim$1$\arcmin\times7\arcmin$ map.  Scheduling constraints are used to ensure
that these strips be oriented as closely as possible to the IRS spectral maps,
so we will cover the same regions in the galaxies at 14--40~\micron\ and
52--99~\micron.
We used the radial profiles of nearby galaxies discussed above to estimate the
typical surface brightnesses at R = 2--4 \arcmin, and computed the 
integration times required.  The fainter regions correspond to 
$\sim$1.2~MJy~sr$^{-1}$,
which implies a $1\sigma$ sensitivity of
4.5~mJy on a point source to achieve $3\sigma$ sensitivity at that surface
brightness.
This is achieved in 4 cycles of 10~s integrations, so we will perform two
cycles at each map position so that the half-slit-width overlaps provide the
total integration time required.

\subsection{IRS Spectroscopy}

The IRS observations consist of: (1) low-resolution radial strip
maps of all 75 galaxies in the long-low channel (14.2--40.0\,\um);
(2) targeted spectroscopy of the centers of all 75 galaxies and
75 extranuclear sources in a subset of galaxies, at 5.3--14.2\,\um\
(low resolution), and 10.0--37.0\,\um\ (high resolution).  

\medskip

\subsubsection{14--40 \um\ IRS Radial Strip Spectroscopy}

We will obtain 0\farcm.9-wide spectral maps for all galaxies from 14.2 to
40.0\,\um\ using the IRS low resolution mode, maximally overlapping the
MIPS SED strip.  These IRS strips will extend radially to about
0.55$R_{25}$, where the average surface brightness is $f_\nu(15 \mu m)
\sim$1~MJy~sr$^{-1}$ (Dale et al.\ 2000).  To minimize the effects of
cosmic rays and bad pixels, we will spatially quadruple-sample by
executing half-slit (length and width) steps parallel and perpendicular to
the long-low slits.  Since the IRS short-low subslits are about one arcmin
long and are almost perpendicular to the long-low slits, we will
complement the 1\arcmin-wide long-low strip maps with short-low maps
(\S~\ref{sec:targeted_low_res}), to obtain complete wavelength coverage
from 5.3 to 40.0 \um\ in the nuclear regions.
Again using the $IRAS$+\ISO-based spectral model of Dale et al.\ (2001),
we find that exposure times of 30 seconds give S/N $\ge 5$ for a
15~\um\ surface brightness of $\sim$1~MJy~sr$^{-1}$, for the cirrus-like SEDs
in the outer disks.

\medskip

\subsubsection{Targeted Low-Res Spectra (5--14 \um)}
\label{sec:targeted_low_res}

Low-resolution spectral maps from 5.3--14.2\,\um, Nyquist-sampled over a
0\farcm3\,$\times$\,0\farcm9 region, will be obtained for the centers of each
galaxy and the 75 extranuclear regions.  This region of the spectrum
is especially valuable for measuring the aromatic dust emission bands
from small grains (including PAH species).  Since the galaxy centers will
also be covered by the 14.2--40.0\,\um\ and 52--99\,\um\ SED strips, we will
have nearly complete spectral coverage of these regions.  Moreover, the
55\arcsec\ sub-slit length will enable useful observations of extended
sources, and the serendipitous regions simultaneously covered by the other
sub-slit will provide data on the local background and quiescent disk
regions.  We will use exposures of 14\,s for the nuclei and 60\,s for
extranuclear regions.  Based on the range of observed mid-infrared surface
brightnesses for the \ISO\ Key Project galaxies ($\sim 10 -
300$~MJy~sr$^{-1}$), we expect S/N $\ge$ 5 in the core of each target.

\medskip

\subsubsection{Targeted High-Resolution Spectra (10--37 \um)}

As for the low-resolution strip maps, we will map targeted regions with
high spectral resolution, using a 5\,$\times$\,3 grid
pattern of half-slit length and half-slit width steps, as shown
in the righthand panel in Figure~7.  Most of our
targets are spatially extended, with significant emission structure over
scales comparable to the sizes of the IRS short-high aperture.  As a
result it would be very difficult to obtain accurate spectrophotometry
across the entire wavelength range (10--37\,\micron) with a single 
pointing observation, because the 
size of the point-spread-function (PSF) varies by so much 
over this wide spectral range.  Our spectral mapping strategy
will provide: 1) adequate spatial coverage to obtain accurate line ratios over
the same physical region in the sources; 2) S/N $\ge 5$ in the principal
diagnostic lines (below); 3) flexible scheduling and roll angle.  

High-resolution observations with the IRS spectrograph are covered in two
echelle modes, a ``short-high" mode at 10--19.5\,\micron\ and a 
``long-high" mode at 19.3--37\,\micron.  Our short-high spectral maps
will cover a total area of 15\farcs7\,$\times$\,23\farcs6,
while the long-high maps cover an area of 33\farcs5\,$\times$\,44\farcs8.
The larger area in the long-high observation is 
necessitated by the larger PSF at the longer wavelengths.  Our basic unit
of coverage for the nuclei and HII regions is well-matched to the area
covered by the short-high slit.
We expect to achieve 1~$\sigma$ noise levels for unresolved emission lines
of approximately $1-2 \times 10^{-18}$~W~m$^{-2}$ in the short-high and
long-high data, except for the region beyond about 32 \um\ which may be a
factor of 2--4 noisier.  Thus, emission lines at the $10-20 \times
10^{-18}$~W~m$^{-2}$ level, similar to what we expect for [NeII] and
[SIII], should be detected at S/N $\ge$ 5--10 over the entire mapped area.
Higher S/N is expected in the centers of the mapped regions, where we
benefit from aperture oversampling.

\subsection{Summary of SIRTF Observations}

The total SINGS observing program is 512 hr, divided roughly equally
between IRAC and MIPS imaging, IRS and MIPS low-resolution spectral
mapping, and the targeted IRS spectroscopy (Table~4).  It is important to bear
in mind that all of the Legacy observing plans will be re-evaluated
after the launch and commissioning phase of SIRTF, and are subject
to revision should there be significant variances in the flight performance
of one or more of the instruments.  The SINGS observing strategy
was deliberately designed to be robust against modest variances in
instrumental performance, so unless the instrumental sensitivities
or capabilities changes substantially, the 
program will still be carried out as described above.  

For a subset of IR-faint sources, including most of the E--S0
galaxies and some of the dwarf irregular galaxies, we have deferred
scheduling the spectroscopic observations until after the IRAC and
MIPS imaging have been obtained, so we can confirm the detection of
a minimum threshold infrared surface brightness to make spectroscopy
feasible.  Detailed information
can be found on the SINGS website (URL: http://sings.stsci.edu).

\section{Ancillary Observations and Complementary Data}

The scientific value of the SIRTF data will be enhanced by 
a large set of ancillary multi-wavelength observations that
are being obtained and assembled for the SINGS galaxies.  Many of 
these data are being obtained by the SINGS team, and will be
archived together with the SIRTF data products (\S\,6).
Our project also has stimulated other groups to undertake 
``complementary" surveys of the SINGS sample, and most of those data 
will be placed in publicly-accessible data archives after
the surveys are completed.  To date supporting data have been obtained
on 20 telescopes, and when these surveys are completed they will
comprise the most comprehensive multi-wavelength dataset ever assembled
for a nearby galaxy sample.  

Table 5 gives a compact summary of these ancillary and complementary
datasets, including information on the field coverages, spatial
resolutions, and approximate numbers of galaxies observed.  A more detailed
galaxy-by-galaxy listing of ancillary and complementary observations
will be maintained on the
SINGS website and updated on a regular basis.  In the remainder of this section 
we briefly summarize each of these datasets.

\subsection{$BVRIJHK$ and \Ha\ Imaging}  

Broadband images in the visible and near-infrared provide an essential
complement to the SIRTF data, providing information on the local 
stellar spectral energy distributions at shorter wavelengths, 
and constraints on the dust extinction, stellar mass distributions, 
and stellar populations heating the dust.  Emission-line images
in \Ha\ provide a local measurement of the stellar ionizing flux
and SFR (Kennicutt 1983), and when combined with the broadband data
can constrain the ages and other properties of the star-forming regions.

Thanks to the generous allocation of observing time from NOAO, matched-field
images in $BVRI$ and \halpha\ have been obtained for nearly all of the 
SINGS galaxies using the KPNO 2.1\,m telescope with CFIM imager, and
the CTIO 1.5\,m telescope with Cassegrain CCD imager.  These imagers
provide field coverage of 10\farcm4 and 8\arcmin\ respectively, and
for larger galaxies a series of positions were observed in order to
provide full coverage to R $\le$ R$_{25}$.  For a handful of
galaxies \halpha\ images are being obtained with the 2K CCD imager on 
the Steward Observatory Bok 2.3\,m telescope.  All data have been
obtained during photometric conditions, or under thin clouds with
separate calibration exposures obtained in photometric weather.

Two sets of near-infrared imaging ($JHK$) of the SINGS galaxies are
being collected.  A set of $JHK_s$ images will be extracted from the 
Two-Micron All Sky Survey (2MASS).  These images will provide complete
and relatively uniform quality near-infrared imaging for the entire
sample, though the shallow depth of the data will limit applications
for spatially-resolved studies (cf. Jarrett et al.\ 2003).  Deeper
imaging is being obtained using the OSIRIS imager on the CTIO 1.5\m 
telescope (southern galaxies), and on the Steward
Observatory 1.5\m (Bigelow) and 2.3\m (Bok) telescopes, using either
the 256$\times$256 infrared camera or the PISCES wide-field
camera (McCarthy et al.\ 2001).  These cameras provide fields of view
of 2\farcm5 -- 3\farcm8 and 8\farcm5 respectively.  Since the
observing time required to obtain high signal/noise at R$_{25}$
would be prohibitive at these wavelengths, we have chosen instead
to obtain high-quality imaging of the central regions of the 
galaxies, where extinction is most problematic at shorter wavelengths.
These images reach a typical depth of $\mu_J$ = 22 mag~arcsec$^{-2}$
or $\mu_{Ks}$ = 21 mag~arcsec$^{-2}$.  

\subsection{HST NICMOS Paschen-$\alpha$ and H-Band Imaging}

In order to provide more robust measurements of the extinction,
massive star formation, and stellar mass distributions in the centers
of the SINGS galaxies, we are obtaining P$\alpha$ emission-line and
$H$-band imaging, using the NICMOS camera on HST.  
Data for appoximately 30 SINGS galaxies are being obtained through
an ongoing HST Snapshot program (GO-9360).  Since Snapshot targets
are observed on an available opportunity basis, the exact number of
galaxies to be observed (and their identity) is uncertain at this time.
In addition, archival observations are available for 19 galaxies,
and we will process these data and deliver them as SINGS data products.
Appropriate redshifted Pa$\alpha$ filters are only available for 
part of the SINGS sample, so coverage of the sample will be incomplete,
but will include a wide enough range of properties to enable a wide
range of scientific applications.  

The images are being obtained with the NIC3 camera, which provides
a field size of 55\arcsec\ square.  For typical exposure times and
line/continuum ratios the Pa$\alpha$ maps will reach a limiting
surface brightness of 
$1.5 \times 10^{-15}$ ergs~cm$^{-2}$~s$^{-1}$~arcsec$^{-2}$
over an area of 0\farcs6 $\times$ 0\farcs6 (for $S/N = 10$).  
The corresponding depth of the $H$-band broad band images will be
19.7 $H$~mag~arcsec$^{-2}$.

The P$\alpha$ recombination
line luminosity is directly tied to the ionizing luminosity in the 
same way as \Ha, but due to its longer wavelength (1.87 \micron) it is
much less susceptible to dust.  In addition the P$\alpha$/\Ha\ ratio
provides a powerful extinction tracer in HII regions (e.g., 
B\"oker et al. 1999; Scoville 
et al.\ 2001; Quillen \& Yukita 2001), and it will provide a valuable
independent constraint on the dust extinction and heating models
in the central regions of these galaxies.

\subsection{Optical Spectrophotometry}

As an aid to the interpretation of the SINGS 10--40\,\micron\ and
52--99\,\micron\ maps, our group is obtaining spectral drift-scans of
the same regions in the 3600--7000 \AA\ region, using Steward Observatory
Bok 2.3\,m telescope for the northern galaxies and the CTIO 1.5\,m telescope
for southern galaxies.  These data are obtained by trailing the galaxy
image back and forth across the slit during an integration, so 
an integrated spectrum of a large rectangular strip in the sky can be
obtained, while preserving the spectral resolution of the narrow-slit
observation.  For each galaxy three one-dimensional integrated spectra 
are obtained.  One spectrum covers the same area as the IRS and MIPS SED radial
strips (\S\,4.4), and a second 
is integrated over the central 20\arcsec\,$\times$\,20\arcsec\
region, and is designed to provide matching spectrophotometry for the
circumnuclear IRS spectra (\S\,4.4).  A third, narrow-slit pointed 
spectrum of the nucleus itself will also be provided. 

The reduced, flux-calibrated spectra will have spectral resolution of 
$\sim$7--8 \AA, and will provide information on the primary diagnostic
nebular emission lines (\Ha, H$\beta$, [OII]$\lambda$3727, 
[OIII]$\lambda$4959,5007, [NII]$\lambda$6548,6583, [SII]$\lambda$6717,6731),
in some cases fainter lines such as [OIII]$\lambda$4363, and 
information on the shape and principle absorption lines in the stellar
spectra.  Note that since the spectra are taken in a drift-scanned mode
there is no point-by-point spatial information.  
Optical spectra for most of the extra-nuclear HII regions in the SINGS sample
(Table 3) are published in the literature, so new observations are not
required.  

\subsection{H$\alpha$ Kinematics}

A group from the Universite de Montreal and the Laboratoire
d 'Astrophysique de Marseille is obtaining \halpha\ kinematics for the SINGS
sample.  The observations will help to delineate the role of gas kinematics
in regulating the SFR, and will provide a fundamental local reference sample 
for 3D kinematics studies
of high-redshift galaxies. All the data are being obtained with a new
GaAs Photon Counting System (Gach et al. 2002). The observations are
obtained on the CFHT 3.6\,m, ESO La Silla 3.6\,m, Observatoire de
Haute Provence (OHP) 1.93\,m, and Observatoire de Mont Megantic (OMM)
1.6\,m telescopes, with fields of view of 3\farcm\ to 12\arcmin.

These data will complement the lower-resolution gas kinematic data being
provided by the HI and CO surveys (below), by providing high spatial
resolution data for the inner parts of galaxies.  These are crucial for
constraining the dark matter density profiles of galaxies
(e.g., Blais-Ouellette et al. 1999).
Such a database will also make it possible to study how the parameters of the
dark matter distribution vary as a function of morphological type, and test
for influences on the star formation properties and histories of the galaxies.

\subsection{UV Imaging and Spectrophotometry}

Imaging and spectra of the SINGS galaxies in the ultraviolet 
(1300--2800\,\AA) provide an especially powerful complement to infrared
data, because this wavelength region directly traces the photospheric
emission of the dominant star-forming and dust-heating population in
most galaxies.  The recently launched Galaxy Evolution Explorer Mission 
(GALEX) is designed to perform
an all-sky survey in two bands over 1350--3000\,\AA, with
angular resolution of $\le$ 5\arcsec\ (Milliard et al.\ 2001).
This is well matched to the resolution of SIRTF, and will be adequate to
quantify the integrated UV emission from the entire sample.  Since
the approval of the SINGS project the GALEX team also has tentatively
planned much deeper imaging of each SINGS galaxy ($\sim$1\,h each at 
1350--1800\,\AA\ and 1800--3000\,\AA).  These deeper images will enable  
point-by-point mapping of the SEDs and SFRs in the galaxies.
All of these data will be released into the GALEX public archive
after a nominal proprietary period, and the images of the SINGS galaxies
will be linked to our archive shortly thereafter.

Images in this wavelength region (1500--2800\,\AA) are also available for
27 of the galaxies from the Ultraviolet Imaging Telescope (UIT)
and/or HST, and IUE spectra of the central 10\arcsec\,$\times$\,20\arcsec\
regions are available for 25 galaxies.  These data will be integrated
into the SINGS database and included in the data archive as well.

\subsection{CO, HI, and Radio Continuum Maps}

Maps of CO and HI emission are important for tracing the cold, gaseous
ISM, and aperture synthesis maps enable a comparison with SIRTF data
at the highest resolution possible.  In addition, measures of radio
continuum emission ($\lambda$ = 20\,cm) provide an alternate tracer of
star formation as well as a means to examine the correlation of far-infrared
and non-thermal radio continuum emission (e.g., Condon 1992; Marsh \& Helou 
1995; Sullivan et al.\ 2001).  

For a core
set of 20 spiral galaxies both CO and HI aperture synthesis observations 
have been acquired as part of the
BIMA Survey of Nearby Galaxies (BIMA SONG), and we
will include the reduced maps in the  SINGS data products
(Regan et al.\ 2001; Helfer et al.\ 2003).  The typical half-power beamwidth
(HPBW) for these maps is 6\,\arcsec, and single-dish total-power measurements
have been incorporated into the reductions to preserve extended emission.
Further BIMA
observations are planned for other northern SINGS galaxies as
motivated by our SIRTF observations.  
Published single-dish CO measurements from a variety of sources 
are available for 36 of the SINGS galaxies (see Table 2).  In addition
the Five College Radio Astronomy Observatory (M.\ Heyer, PI) is 
performing an On-The-Fly mapping survey of a broader subset of SINGS galaxies.
Observations are being obtained using the FCRAO 14\,m telescope 
with the SEQUOUA focal plane array (16 pixels for 
extra-galactic observations).  Typical maps consist of 5$\times$5 to 
8$\times$8 pointings, with a target sensitivity 
is 5 milliKelvin in 5 MHz wide channels, chosen to match the sensitivity 
of the 12\,m telescope observations from BIMA SONG.

We are assembling a
database of 30--45\,\arcsec resolution HI maps for northern SINGS
galaxies.  These data come from the dedicated Westerbork (WSRT) HI
survey, WHISP (Swaters et al. 2002; Swaters \& Balcells 2002), as well
as new and archival VLA observations.  In addition the nearest SINGS galaxies 
(d $<$ 10 Mpc)  are being observed in HI at high spatial and spectral
resolution mapping as part of a VLA Large Proposal (F. Walter, PI).
This will provide data for $\sim$35 galaxies of the SINGS sample 
at high spatial resolutions of 7\arcsec\ and velocity resolution
2.5--5\,km\,s$^{-1}$. These data will be used to
investigate the small scale structure of the atomic ISM and to obtain
spatially resolved velocity dispersion maps for each galaxy.  The
field of view of all HI observations is $\sim$30\arcmin, and nicely
encompasses even the largest galaxies of the SINGS sample.

We are also assembling a database of 30--45\,\arcsec\ resolution maps 
of the 20-cm 
radio continuum emission from SINGS galaxies.  These data come from the 
VLA archives, as well as new observations of 40 SINGS galaxies, and seek 
to achieve an rms noise level half that of the NRAO VLA Sky Survey
(0.2--0.3 mJy\,beam$^{-1}$; Condon et al.\ 1998) at 
equal or higher resolution.  
In a parallel effort a group led by Robert Braun (NFRA) is using
the Westerbork Synthesis Radiotelescope (WSRT) to obtain deep
20 cm radio continuum and coarse-resolution  HI line maps for 
30 large (D$_{25} > 5$\arcmin) northern SINGS galaxies (see Table 4).
These maps will have a resolution of approximately 12\arcsec\ $\times$
12\arcsec/$\sin \delta$.  The continuum data will include separate
or combined 18~cm and 22~cm images reaching a limit of 15 $\mu$Jy\,beam$^{-1}$
in each band (including spectral index, polarization, and rotation 
measure maps).

\subsection{Submillimeter and ISO Observations}

As can be seen from Figures 1 and 8, the thermal dust emission of
galaxies usually extends to wavelengths well beyond 160~\um,
the longest wavelength for SIRTF instruments.  This emission can
represent a significant fraction of the total dust luminosity, and
is critical for constraining the temperatures and masses
of the cold dust component in the ISM (e.g., Siebenmorgen, Kr\"ugel,
\& Chini 1999; Bendo et al.\ 2003).

To properly trace continuum dust emission from the far-infrared into the 
submillimeter, we have acquired data from the James Clerck Maxwell Telescope 
(JCMT) archives maintained by the Canadian Astronomy Data 
Centre\footnote{http://cadcwww.hia.nrc.ca/}.
At the time of this writing, the archive contain Submillimetre Common-User
Bolometer Array (Holland et al.\ 1999) data for 24 SINGS galaxies. 
Only 17 galaxies are detected in at
least the 850\,\micron\ waveband, although many of these galaxies are also 
detected at 450\,\micron. For some galaxies, the data include stare photometry 
and jiggle maps of the inner 2.3\arcmin, whereas for other galaxies, the
data include composite jiggle maps and scan maps 
that may extend beyond 10\arcmin\ apertures.  The data are being 
processed using the SURF data reduction package (Jenness \& Lightfoot 1998)
following standard data reduction procudures.  Additional submillimeter
observations are planned to acquire data for at least the brightest infrared
sources in the sample.

Data for 46 of the 75 SINGS galaxies are available from the Infrared
Space Observatory (Kessler et al.\ 1996) 
archives\footnote{http://www.iso.vilspa.esa.es/ida/index.html}.  These data
come from all four of the ISO instruments, which we may use to
supplement our existing data and check our flux calibration.  
The ISOCAM (Cesarsky et al.\ 1996) and ISOPHOT (Lemke et al.\ 1996) 
broad band data may be useful for adding photometry to the observed 
spectral energy distributions of the SINGS galaxies,  
for checking the calibration of IRS and MIPS data, and extending the
SED coverage redward to 240~\um.  ISOSWS (de Graauw et al.\
1996) data provides spectra with higher spectral resolution than IRS 
that can be used to closely examine the wavelength profiles of emission line 
features. Finally, ISOLWS (Clegg et al.\ 1996) spectral data, which spans 
wavelengths longer than what IRS can observe, can provide 
additional diagnostic spectral lines for
modeling star formation and the interstellar medium.

\subsection{X-Ray Maps}

By the end of 2003, 35 galaxies (about 50\%) of the SINGS sample will have
been observed with the Chandra X--ray observatory (with integration times
ranging from 2 to 70\,ksec). We anticipate that more SINGS targets will be
added to the list in the years to come.  As part of SINGS, processed 
X--ray maps for
these galaxies will be made available in various energy bands (both photon
and adaptively smoothed images). Chandra's arcsecond resolution will be
essential for a detailed comparison of the X--ray data with observations
obtained at other wavelengths and for isolating point sources from
the underlaying diffuse hot gas. 
These X-ray point sources (such as X--ray binaries and supernova
remnants) provide important clues to the heating mechanisms of the
ambient medium. In complementary fashion, the multi--wavelength SINGS
data will be important to constrain the nature of the Chandra point
sources. The Chandra data will also allow us to derive physical
properties (such as tempertature, density, pressure) of the hot phase
of the ISM which will be visible as diffuse, soft X--ray emission.
The SINGS team is also collaborating on proposals to enlarge
the subset of galaxies with high-quality Chandra maps.

\section{SINGS Data Processing, Data Products, and Website}

As with all Legacy programs, the SIRTF observations are completely
non-proprietary.  This means that the pipeline-calibrated data,
including calibrated single-pointing images (and rudimentary mosaic images)
and individual spectra will be available to the public immediately.
These data will enable a wide range of scientific investigations 
in their own right, and provide invaluable samples of flight data
for planning of future GO observing programs.  

A primary scientific output of the SINGS project will be added-value
data products, including high-fidelity image mosaics, enhanced-resolution 
dithered images, spatially co-aligned multi-band images, one-dimensional 
spectral extractions, and full 3D spectroscopic data cubes.  A typical
image or spectroscopic observation of a SINGS galaxy is built from 
15--815 individual SIRTF datasets, so these data products will be the
primary source of science-grade data for many users.  We also will deliver
user versions of tools for extracting
source spectral energy distributions from multi-wavelength sets of images
and for building and analyzing spectral data cubes from sets of multiple
IRS pointings.


\subsection{IRAC and MIPS Data Products}

We will produce and deliver two sets of IRAC data products:
1) image mosaics in native IRAC resolution;
2) enhanced mosaics.
The first set of images will possess the native resolution of the 
instrument, which is limited by the pixel size of the IRAC detectors
(1\farcs2).  Since this pixel size undersamples the telescope's point 
spread function at all wavelengths, we plan to construct a second 
set of mosaic images that are processed using 
the drizzle technique (Fruchter \& Hook 2002) to reconstruct the point 
spread function, and thus provide higher spatial resolution images.  

The processing and production of the mosaic images will also include
a number of other checks and enhancements to the basic calibrated
data.  Cross-correlation of the individual pointings will be used
to derive spatial offsets between subexposures, and check the pointing
reconstruction given in the image headers.  We will also check for 
and if needed remove other instrumental signatures such as detector
bias level drifts, residual cosmic rays, and residual hot pixels.
Asteroids will be detected by combining the time-separated 
exposures that are being designed for this purpose.

The MIPS data products will include 24, 70, and 160~\micron\ mosaicked
images and a SED strip spectral cube for each of the 75 SINGS galaxies.
The initial processing will proceed in similar fashion to IRAC,
with a combination of manual checks on data quality followed
by a carefully supervised processing of individual observations to
create mosaicked images. 
The MIPS image data products will be delivered in two versions.  The first
version will encompass the entire area observed at each wavelength.
Due to the placement of the three focal plane arrays and their
different sizes, this will result in different fields-of-view at each
wavelength for both photometry and scan mode observations.  The second
version will have images which are background-subtracted (constant
value), share a common field-of-view, and have been regridded to a
common pixel scale.  

The MIPS SED data will be processed
in the same way, before being combined with the \cbsm\ package (\S\,6.3.2) to
create spectral data cubes from individual mosaicked SED pointings.
The resulting data products will be delivered in two versions as well.  
The first will consist of two spectral cubes, one for the object (without
background removal) and one for the nearby background observations.
The second version will be a single spectral cube with the nearby
background observations subtracted from the object observations.

A program for automated mosaicking of MIPS images
(named mips\_enhancer) 
is being developed by the MIPS Instrument Team in collaboration with 
the SINGS team.  The goal is to use the redundancy inherent in the MIPS
observational strategy to remove residual instrumental signatures in
the 70 and 160 \micron\ data.  
The program will be continually improved as knowledge of the detector
instrumental signatures increases, and a final delivery of mosaicked images
and spectra that incorporate our accumulated knowledge to date will
be made at the end of the project.


\subsection{IRS Data Products and Tools}

The SINGS IRS data products consist of three separate types for each
mapped region:  1) full, 3D spectral data cubes;  2) selected line
and continuum spectral maps; and 3) representative one-dimensional
extracted spectra.  
Tools for creating and visualizing the spectral cubes will also be
available, and these are described below (\S 6.3.2).  

\subsubsection{IRS Spectral Mapping Cubes}
\label{sec:spectral-cubes}
The fundamental data product generated from the IRS spectral mapping
observations is a spectral cube, created by combining spatially
overlapping data from different slit pointings into a three dimensional
data cube with coordinate axes $(\alpha,\delta,\lambda)$.  
The other IRS data products are derived directly from the spectral cube.
A single spectral cube will be created for each IRS sub-slit (6 in
total) used in all nuclear, extra-nuclear and radial strip observations.

The cubes are constructed using a flux-conservative algorithm somewhat
related to drizzle (Fruchter \& Hook 2002), but operating
in three dimensions on collections of 2D spectral images.  Spatial and
spectral smoothing, and other error-correlating operations are avoided
by mapping pixels from the raw spectra directly to the spectral cube.
Individual single-wavelength spatial samples, consisting of tilted
``pseudo-rectangles'' approximately as long as the slit, and with widths
equal to the wavelength sampling interval, are tiled on the detector,
and clipped against its pixel grid.  The resulting partial pixels are
rotated to match the ``sky grid'' of the final cube, both to remove any
intrinsic instrumental spectral line tilt, and to reference all
observations to a common position angle of the slit on the sky.  Samples
collected from different spectral orders which overlap in wavelength,
and from adjacent spectrograms within the mapping observation set are
combined together in the resultant cube.

The entire set of rotated partial pixels is clipped again against the
output sky grid, which is constructed with an adjustable pixel spacing
defaulting to the native plate scale of the IRS module in use (none of
the IRS modules is significantly undersampled).  The clipped area for
each pixel portion is retained, and in this way an area-conserving
mapping of spectrogram pixels to cube pixels is built up, and used to
construct the final cube.  Area/error weighting is the default method
for combining the contributing pixels, but trimmed weighted means are
also possible.

\subsubsection{IRS Spectral Maps and 1D Spectra}

Spectral maps in [Ne\,II] (12.8\,$\mu$m, $\phi$=22\,eV), [Ne\,III]
(15.5\,$\mu$m, $\phi$=41\,eV), and [Si\,II] (34.8\,$\mu$m, $\phi$=8\,eV)
line features will be produced from all the SINGS nuclei and targeted
HII region cubes.  These maps will be generated from the spectral cubes
based on simple fits to the underlying continuum.  The
continuum-subtracted maps, along with the derived continuum maps used in
the subtraction will be included.

In addition to the Ne and Si atomic line maps, we will produce maps in
the combined [7.7\,$\mu$m + 8.6\,$\mu$m], and 11.3\,$\mu$m PAH features
for all the SINGS nuclei and targeted HII regions.  Since these broad
dust features dominate the spectrum of star-forming galaxies at these
wavelengths, we will not perform continuum subtraction.
Instead, we will extract a set of line-free continuum maps that 
will provide information on the underlying continuum, as well as
providing additional information on the galaxy SED at wavelengths
that fall between the IRAC and MIPS filters (14--40 \micron).

By selecting spatial regions in the maps or cubes we can create 1D
spectral extractions for a given IRS sub-slit.  We will produce 1D
spectra for all SINGS nuclei and mapped HII regions.  These ``source"
spectra will be extracted from a suitable region (matched to the PSF),
centered on the map peak at either 12 or 25\,$\mu$m, depending upon the
IRS module used to create the map.  The surrounding areas in the map
will be used to create a local ``background" 1D spectrum, which will
suitably scaled and subtracted from the source spectrum.  Both the
background-subtracted source spectrum and the background spectrum itself
will be included in the data products.

\subsection{User Software Tools}

Most of the SINGS data products will be in the form of image mosaics
and extracted spectra which can be readily analyzed by the generic
SIRTF data analysis programs or by standard image analysis packages
such as IRAF or IDL.  However our team is developing two software
packages that will be specifically tailored to analysis of the SINGS
data.  One of these tools ({\textsc SEDTIAP}) will enable the 
batch-mode generation of PSF-matched SEDs from a set of individual
multi-wavelength images.  The \cbsm\ tool will generate 3D spectral
data cubes from IRS mapping observations, and provide a variety of
user software for analyzing the spectral data cubes.  

\subsubsection{SED Tool and Image Analysis Program}

The {\textsc SEDTIAP} program is an IDL-based tool that 
rotates, rescales, interpolates/convolves the input image
data to a common astrometric coordinate system and resolution, using
the user-selected PSF from a set of available choices and assuming
bi-linear interpolation with flux conservation to resample the
images. The co-aligned images are displayed for the user to check
alignment and a data-cube containing the processed images is saved to
disk. Apertures for both sources and background are specified
interactively by the user, and the total and background-subtracted
fluxes for all the images in the data-cube are saved in a file (the
SED file).


\subsubsection{\cbsm}
\label{sec:cubism}

Spectral cubes are constructed from the IRS spectral mapping data sets
using a custom software package, \cbsm.  The cubes themselves and the
algorithms used to create them are discussed in
\S\,\ref{sec:spectral-cubes}.  \cbsm\ is written in IDL, and
consists of four main components:  
 i) the cube-builder backend (see \S\,\ref{sec:spectral-cubes});
 ii) CubeProject, a graphical tool for organizing and querying cube
  inputs and driving the cube creation; 
 iii)  CubeView, a visualizer for both 2D spectrograms and spectral
  cubes, facilitating 1D extractions;
 iv)  CubeSpec, for visualizing 1D extracted cube spectra, and
  generating stacked maps.

CubeProject provides the interface from which all cube-building
operations are performed.  It encapsulates individual data records into
``projects'', each of which includes all the calibration information and
data required to build a single cube for one IRS sub-slit.  
Individual data records can be added, removed, disabled, queried for
header information, and sorted in a variety of ways.  Visual feedback of
the pixel clipping process is provided, and the individual spectrograms
or the final cube can be sent to CubeView for display.


CubeView permits both individual spectral data records and the full cube
to be displayed and analyzed.  A suite of traditional viewer tools
(zooming, colormap and histogram scaling, box statistics, photometry,
etc.) is available.  In addition, various specialized tools are
available for reporting wavelength and order (for spectrograms) or
celestial position (for spectral cubes) within the image, selecting and
setting the apertures which delineate the order data to be included in
the cube build, navigating through wavelength planes, and extracting
spectra from selected regions.
In the latter case, the extracted 1D spectrum is displayed in a related
tool, CubeSpec, which can be used to visit individual planes of the cube
and to create line or feature maps.  Arbitrary peak and continuum
wavelength regions can be defined on the spectrum, and the resultant
maps are displayed in CubeView, where they can be saved or exported.
Stored region sets for important lines and features in the wavelength
range are available, including 8 and 24\,$\mu$m sets corresponding to
the appropriate IRAC and MIPS bandpasses, to enable easy
cross-calibration between instruments.  The continuum can also be fit
and removed using the defined background regions.

\subsection{Ancillary and Complementary Data Products}

\begin{figure*}
\epsscale{1.5}
\plotone{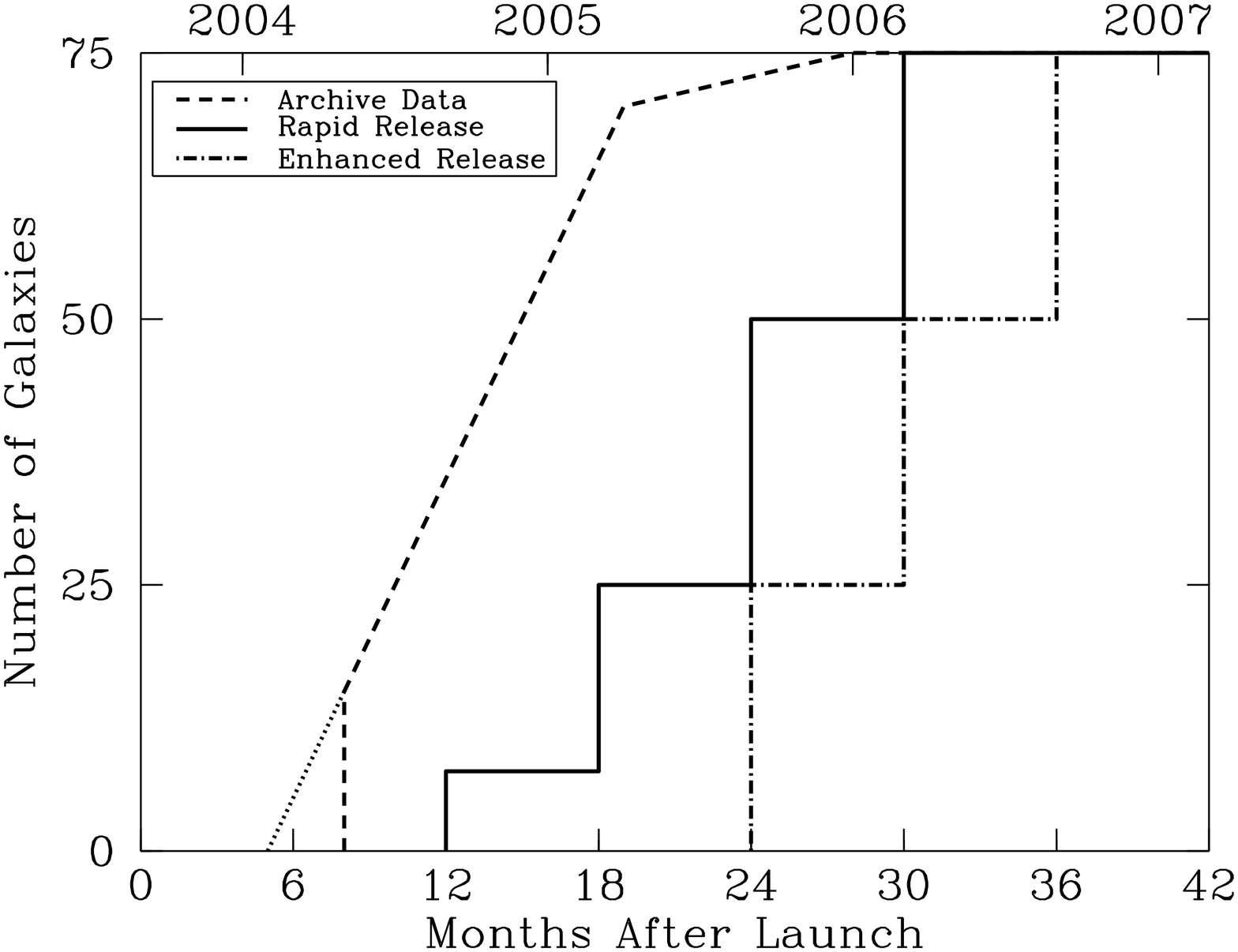}
\caption{This graph shows the anticipated schedule for availability
of SINGS data in the IRSA archive.  The observing and data release
schedules are tied to time after the SIRTF launch, as indicated on
the bottom axis.  An approximate corresponding calendar schedule
is shown on the top axis, for an assumed launch date of 1 Sept 2003.
From left to right, the dashed line shows the approximate flow of
pipeline-processed observations into the SIRTF archive.  The solid
line shows the anticipated schedule for delivery of SINGS rapid release
reduced data products (see Table~5) and the corresponding ancillary
data.  The dot-dashed line shows
the corresponding delivery of SINGS enhanced data products.} 
\end{figure*}

As discussed in \S\,5, the scientific value of the SIRTF data will be enhanced
substantially by the addition of images and spectra of the galaxies
at other wavelengths.  Some of these data (e.g., visible and near-infrared
imaging, visible spectra) are being obtained by the
SINGS team and will be delivered to the NASA/IPAC Infrared Space Archive
(IRSA) as formal ancillary data products.  They will be fully
reduced and will include headers with full world coordinate system
(WCS) information, flux calibration
information, and when relevant comments on the photometric and/or
astrometric accuracy of the images.  Data from other sources that are 
re-reduced by our team (e.g., submillimeter, X-ray images)
will also carry comments or documentation on the reduction procedures
and any qualifications on the data quality.  Complementary data from other
groups (Table~5) will be posted either in the SINGS archive or on
independent websites, according to the wishes of the groups
involved.  Information and links to these sites will be provided on the
main SINGS website (below).  

\subsection{Data Archive and The SINGS Website}

All public access to SINGS data will be made through the SIRTF
Legacy archive at IRSA.\footnote{ 
URL: http://irsa.ipac.caltech.edu}\  Data releases and access will
be administered by the SIRTF Science Center, and prospective
users should follow announcements from SSC to obtain more updated
information.  
In addition the SINGS team maintains its own public website.\footnote{
URL: http://sings.stsci.edu}\  The site contains detailed information
on the project and galaxy sample, and will be regularly updated to 
provide information on observations and data products.  

Scheduling of the SIRTF observations and the subsequent data 
releases is subject to the launch date, and scheduling of
the various GTO, Legacy, and GO observations thereafter.  However
the graph in Figure 9 provides an approximate projection of the
expected data flow, as a function of month after launch.  
We expect approximately 5 galaxies per month to be
observed on average, beginning approximately 3--4 months after
the launch of SIRTF.  The first of these data will be released
to the public with the opening of the SIRTF archive, approximately
8 months after launch, and after that time the basic calibrated 
data and other pipeline observations will be posted in the archive
shortly after the observations are taken (as indicated by the 
leftmost dashed line in Figure 9.  A small fraction of the spectroscopic
observations (galaxies with questionable detections, IR-selected
extranuclear targets) will not be scheduled until IRAC and MIPS images
for those galaxies are in hand, and this is indicated by the turnover
in the dashed line at the top of the figure.

In order to expedite
the release of science-grade products to the user community 
we have defined two levels of product for each instrument,
as summarized in Table~6.  The ``rapid release" products consist
of processed and mosaicked images for IRAC and MIPS, and 1D spectral
extractions and line maps for IRS.  Each product will incorporate
full checking and validation of the pipeline data by members of the 
SINGS team, calibration and astrometric checks and refinements, 
combination of individual observations into mosaic images or spectral
maps, and corrections for instrumental signatures on a best-effort 
basis, using the knowledge we have acquired at the time.  

Approximately 12 months later the SINGS team will deliver a second
generation set of ``enhanced" data products (Table~5).  These will
include improved resolution dithered IRAC mosaic images, MIPS
images which take into account our accumulated experience with removing
detector transient features, full spectral data cubes and the 
associated Cubism and Cubeview tools, and if necessary improved 
calibrations for all relevant data products.  Our current projection 
for these deliveries is shown by the rightmost dot-dashed line in Figure~8.
In order to maximize the scientific value of the SIRTF data products
for a given galaxy, releases will include the relevant ancillary 
and complementary data for that galaxy (as available), to provide a coherent
multi-wavelength dataset.  
It is anticipated that all of the data described in this
paper will be in the public archive by the end of the project 
(approximately 3 years after the SIRTF launch).

\acknowledgements

We are pleased to acknowledge the support of the staff at SSC
and SIRTF, with special thanks to Michael Bicay, Tony Marston, and
our SSC liaison scientist, Nancy Silbermann.  Likewise we thank
the Director's Office and staff at KPNO and CTIO for their generous
award of observing time for ancillary observations, and their capable
support during the observing runs.  We also extend
special thanks to IRS PI Jim Houck and his team for cooperating on
observations of common targets, and to the IRS SMART team, as well
as the rest of the IRS Data Products interest group, for their 
cooperation on the development of IRS reduction software.  We are
also indebted to MIPS PI George Rieke and his team for their close
cooperation and collaboration in the development of mapping and
data processing software for observations of extended objects.  We
are also very grateful to John Moustakas and Eric Murphy for assisting in  
the ancillary observations and observing planning.
Finally we thank Philippe Amram, Robert Braun, Claude Carignan, Mark Heyer, Bob
Joseph, Thijs van der Hulst for leading projects to obtain
additional ancillary data for the SINGS galaxies.  
Support for this work, part of the Space Infrared Telescope Facility 
(SIRTF) Legacy Science Program, was provided by NASA through an award 
issued by the Jet Propulsion Laboratory, California Institute of 
Technology under NASA contract 1407.

\newpage

\begin{deluxetable}{ll}
\tablecaption{Range of Properties in the Galaxy Sample}
\tablewidth{0pc}
\tablecolumns{2}
\tablehead{
\colhead{Property} &
\colhead{Range} \cr
}
\startdata
Hubble Type~~~~ &  E~ ---~ Irr \cr
$M_R$       &  $-$12.5~ ---~ $-$23.5 \cr
L$_V$       &  $5 \times 10^6$~ ---~ $2 \times 10^{11}$ L$_\odot$~~ (0.0003--10 L$^*$) \cr
L(IR)     &  $<$10$^7$~ ---~ $3 \times 10^{11}$ L$_\odot$ \cr
L(IR)/L$_R$  &  $<$0.02~ ---~ 42 \cr
F(60)/F(100) &  0.16~ ---~ 1.2 \cr
O/H (0.4 $R_0$)         &  0.05~ ---~ 3 (O/H)$_\odot$ \cr
M$_{gas}$/M$_{stars}$   &  $<$0.001~ ---~ 5 \cr
SFR          &  0~ ---~ 15 \Ms/yr \cr
SFR/$L_V$    &  $<$10$^{-8}$ ---~ 10$^{-4}$ M$_\odot$/yr/$L_\odot$ \cr
\enddata
\end{deluxetable}

\begin{deluxetable}{lrrrcrrr}
\tabletypesize{\scriptsize}
\tablenum{3}
\tablecaption{SINGS Spectroscopic Sample (Optically Selected Targets)} 
\tablewidth{0pc}
\tablecolumns{8}
\tablehead{
\colhead{Region} & \colhead{RA} & \colhead{Dec} & \colhead{Radius} & 
\colhead{$12+\log(O/H)$} & \colhead{A(V)} & \colhead{$\log$\,L(H$\alpha$)} & 
\colhead{f(Br$\alpha$)} \cr
\colhead{} & \colhead{J2000} & \colhead{J2000} & \colhead{kpc} & 
\colhead{} & \colhead{mag} & \colhead{ergs s$^{-1}$} & 
\colhead{$10^{-18}$ W m$^{-2}$} \cr
\colhead{(1)} & \colhead{(2)} & \colhead{(3)} & \colhead{(4)} & \colhead{(5)} &
\colhead{(6)} & \colhead{(7)} & \colhead{(8)}  \cr
}
\startdata
NGC 628~~H 292  & 01:36:45.10 & 15:47:51  &  4.0  &  9.1 & 2.0 &  40.4 & 45 \cr
NGC 628~~H 572  & 01:36:37.50 & 15:45:12  &  6.8  &  9.0 & 0.9 &  39.5 &  6  \cr
NGC 628~~H 627  & 01:36:38.80 & 15:44:25  &  8.9  &  8.8 & 0.9 &  40.2 & 30  \cr
NGC 628~~H 13   & 01:36:35.50 & 15:50:11  &  11.7 &  8.6 & 1.5 &  40.1 & 20  \cr
&&&&&&&\cr
NGC 2403~~HK 361 &  07:36:46.55 & 65:36:54 &  1.0 &  8.8 & 0.4 & 39.5 & 60  \cr
NGC 2403~~HK 270 &  07:36:53.50 & 65:36:40 &  1.1 &  8.7 & 0.3 & 39.4 & 45 \cr
NGC 2403~~HK 128 &  07:37:08.19 & 65:36:33 &  2.8 &  8.6 & 0.8 & 39.9 & 150 \cr
NGC 2403~~HK  65 &  07:37:19.48 & 65:33:57 &  3.5 &  8.6 & 1.8 & 39.6 & 70  \cr
NGC 2403~~HK 542 &  07:36:21.05 & 65:36:55 &  3.7 &  8.6 & 0.5 & 39.6 & 80  \cr
NGC 2403~~VS 9   &  07:36:29.94 & 65:33:43 &  6.2 &  8.1 & \nodata & 39.0 & 20  \cr
&&&&&&&\cr
Ho II~~HSK 45 & 08:19:13.30 & 70:43:08 & \nodata & 8.5 & \nodata & 39.1 &  22 \cr
Ho II~~HSK 67 & 08:19:27.00 & 70:41:59 & \nodata & 8.6 &  & 38.5 &   5 \cr
Ho II~~HSK 70 & 08:19:28.80 & 70:42:21 & \nodata & \nodata & \nodata & 38.6 &   8 \cr
Ho II~~HSK  7 & 08:18:50.10 & 70:44:48 & \nodata & \nodata & \nodata & 38.5 &   6  \cr
&&&&&&&\cr
M81~~HK 230 & 09:56:00.44&  69:04:02 &   4.1 &   9.2 &   0.4&   38.7&   10 \cr
M81~~HK 343 & 09:55:40.67&  68:59:45 &   4.7 &   8.9 &   0.0&   38.6&    8  \cr
M81~~HK 453 & 09:55:24.43&  69:08:15 &   4.8 &   8.9 &   1.0&   39.0&   20 \cr
M81~~HK 268 & 09:55:53.17&  68:59:04 &   5.1 &   8.8 &   0.7&   39.3&   40 \cr
M81~~HK 652 & 09:54:56.63&  69:08:47 &   5.6 &   8.8 &   0.7&   39.0&   20 \cr
M81~~HK 741 & 09:54:42.26&  69:03:36 &   7.7 &   8.7 &   0.5&   39.1&   25 \cr
M81~~Munch 1 & 09:56:17.42&  68:49:50 &   15.0&    8.1&    0.2&  38.0&   23 \cr
&&&&&&&\cr
NGC 4559~~ZKH 20  & 12:35:56.50 & 27:57:40 &  1.4 &  8.9 & 1.1 & 39.5 &  8\cr
NGC 4559~~ZKH 18  & 12:36:00.80 & 27:56:22 &  4.1 &  8.8 & 0.7 & 39.9 & 18   \cr
NGC 4559~~ZKH 17  & 12:36:02.40 & 27:56:45 &  4.7 &  8.8 & 1.1 & 39.7 & 10  \cr
NGC 4559~~ZKH 4   & 12:35:52.10 & 27:59:10 & 12.3 &  8.6 & 0.4 & 39.3 &  5 \cr
&&&&&&&\cr
NGC 4736~~HK 52  &   12:50:49.64 & 41:07:23 & 0.9 &  9.1 & 0.5 & 39.0 & 10 \cr
NGC 4736~~HK 53  &   12:50:49.64 & 41:07:34 & 1.0 &  9.0 & 1.0 & 39.0 & 10 \cr
NGC 4736~~HK 8/9 &   12:50:56.27 & 41:07:20 & 1.0 &  9.0 & 0.5 & 39.3 & 20 \cr
&&&&&&&\cr
M51~~CCM 107 & 13:29:53.10 & 47:12:40  &  2.5  &  9.4  &  1.7 &  39.9 &  35 \cr
M51~~CCM  72 & 13:29:44.10 & 47:10:21  &  5.0  &  9.3  &  1.9 &  40.3 &  70 \cr
M51~~CCM  71 & 13:29:44.60 & 47:09:55  &  5.3  &  9.3  &  0.9 &  39.8 &  23 \cr
M51~~CCM   1 & 13:29:56.20 & 47:14:07  &  5.8  &  9.2  &  2.6 &  40.1 &  50 \cr
M51~~CCM  10 & 13:29:59.60 & 47:14:01  &  6.4  &  9.2  &  2.2 &  39.8 &  23  \cr
M51~~CM  71A & 13:29:39.50 & 47:08:35  &  9.5  &  9.1  &  2.7 &  40.2 &  59 \cr
&&&&&&&\cr
NGC 6822~~Hub V & 19:44:52.85 & $-$14:43:11 & \nodata &  8.1 & \nodata & 38.3 & 33  \cr
NGC 6822~~Hub X & 19:45:05.24 & $-$14:43:13 & \nodata &  8.1 & \nodata & 38.2 & 75 \cr
NGC 6822~~Hub I & 19:44:31.64 & $-$14:42:01 & \nodata &  8.1 & \nodata & 37.9 & 120 \cr
NGC 6822~~Hub III & 19:44:34.05 & $-$14:42:22 & \nodata &  8.1 & \nodata & 37.5 &  10  \cr
&&&&&&&\cr
NGC 6946~~H 4    & 20:35:16.68 & 60:10:57 & 6.0 &  9.1 &  2.2  & 39.8 & 40 \cr
NGC 6946~~HK 3   & 20:35:25.12 & 60:10:03 & \nodata &  9.0 &  1.5  & 40.0 & 75 \cr
NGC 6946~~HK 288 & 20:34:52.29 & 60:12:41 & \nodata &  9.0 &  1.5: & 39.7 & 35 \cr
NGC 6946~~H 40   & 20:34:19.48 & 60:10:09 & 7.7 &  8.9 &  1.2  & 39.5 & 20 \cr
NGC 6946~~H 28   & 20:34:39.03 & 60:13:35 & 8.6 &  8.9 &  1.2  & 39.5 & 20 \cr
\enddata
\end{deluxetable}

\begin{deluxetable}{lrrllrrrrrrrrrrl}
\tabletypesize{\scriptsize}
\rotate
\tablenum{2}
\tablecaption{SINGS Galaxy Sample}
\tablewidth{0pc}
\tablecolumns{16}
\tablehead{
\colhead{Galaxy} &
\colhead{RA} &
\colhead{Dec} &
\colhead{Type} & 
\colhead{Nuc} & 
\colhead{V$_r$} &
\colhead{Dist} &
\colhead{$D_{25}$} & 
\colhead{M$_{opt}$} & 
\colhead{W$_{20}$} &
\colhead{${{\rm L}_{IR}}\over{{\rm L}_{opt}}$} & 
\colhead{${{\rm F}_{60}}\over{{\rm F}_{100}}$} &
\colhead{$\log\,M_{H{\small I}}$} &
\colhead{$\log\,M_{H_2}$} &
\colhead{SFR} &
\colhead{Group} \cr
\colhead{} & 
\colhead{J2000} & 
\colhead{J2000} & 
\colhead{} &  
\colhead{} & 
\colhead{km/s} &
\colhead{Mpc} &
\colhead{arcmin} & 
\colhead{mag} & 
\colhead{km/s} &
\colhead{} &
\colhead{} & 
\colhead{M$_\odot$} &
\colhead{M$_\odot$} &
\colhead{M$_\odot$/yr} \cr
\colhead{} \cr 
\colhead{(1)}  &  \colhead{(2)}  &  \colhead{(3)}  &  \colhead{(4)}  &  \colhead{(5)}  &
\colhead{(6)}  &  \colhead{(7)}  &  \colhead{(8)}  &  \colhead{(9)}  &  \colhead{(10)} &
\colhead{(11)} &  \colhead{(12)} &  \colhead{(13)} &  \colhead{(14)} &  \colhead{(15)} &  \colhead{(16)} \cr 
}
\startdata
NGC~0024&  00:09:56.7 & -24:57:44 & SAc   & \nodata&    554&  8.2&  5.8$\times$1.3 &  $-$18.4& 222. &       0.20&    0.35&    9.07 &    \nodata &  \nodata &  19 -8 +7   \cr
NGC~0337&  00:59:50.3 & -07:34:44 & SBd   & \nodata&   1650& 24.7&  2.9$\times$1.8 &  $-$20.3& 264. &       1.20&    0.49&    \nodata & \nodata &  4.3     &  61 -23 +21 \cr
NGC~0584&  01:31:20.7 & -06:52:05 & E4    & \nodata&   1854& 27.6&  4.2$\times$2.3 &  $-$22.2& \nodata & \nodata& \nodata&    \nodata & \nodata &  \nodata &  52 -7      \cr
NGC~0628&  01:36:41.8 & +15:47:00 & SAc   & \nodata&    657& 11.4& 10.5$\times$9.5 &  $-$20.9& 74. &        0.72&    0.31&    10.11 &   9.49 &     4.0     &  17 -4      \cr
NGC~0855&  02:14:03.6 & +27:52:38 & E     & \nodata&    610&  9.6&  2.6$\times$1.0 &  $-$17.7& \nodata &    0.23&    0.48&    \nodata & \nodata &  \nodata &  \nodata    \cr
NGC~0925&  02:27:16.9 & +33:34:45 & SABd  & H      &    553& 10.1& 10.5$\times$5.9 &  $-$20.6& 224. &       0.18&    0.29&    9.79 &    $<$9.03 &  2.4     &  17 -1      \cr
NGC~1097&  02:46:19.0 & -30:16:30 & SBb   & L      &   1275& 16.9&  9.3$\times$6.3 &  $-$22.4& 402. &       0.75&    0.40&    10.03 &   \nodata &  4-8:    &  51 -3 +1   \cr
NGC~1266&  03:16:00.7 & -02:25:38 & SB0   & L      &   2194& 31.3&  1.5$\times$1.0 &  $-$21.1& \nodata &    5.49&    0.76&    \nodata & \nodata &  \nodata &  \nodata    \cr
NGC~1291&  03:17:18.6 & -41:06:29 & SBa   & \nodata&    839&  9.7&  9.8$\times$8.1 &  $-$22.0& 84. &        0.02&    0.17&    9.19 &    \nodata &  0.4     &  53 -9 +7   \cr
NGC~1316&  03:22:41.7 & -37:12:30 & SAB0  & L      &   1760& 26.3& 12.0$\times$8.5 &  $-$23.5& \nodata &    0.04&    0.41&    $<$8.87 & \nodata &  \nodata &  51 -1      \cr
NGC~1377&  03:36:39.1 & -20:54:08 & S0    & H      &   1792& 24.4&  1.8$\times$0.9 &  $-$19.6& \nodata &    2.04&    1.23&    \nodata & \nodata &  \nodata &  \nodata    \cr
NGC~1404&  03:38:51.9 & -35:35:37 & E1    & \nodata&   1947& 25.1&  3.3$\times$3.0 &  $-$22.9& \nodata & \nodata& \nodata&    \nodata & \nodata &  \nodata &  51 -1      \cr
NGC~1482&  03:54:39.3 & -20:30:09 & SA0   & \nodata&   1655& 22.0&  2.5$\times$1.4 &  $-$20.5& \nodata &    4.65&    0.77&    $<$8.88 & 9.47 &     1.5     &  51 -4      \cr
NGC~1512&  04:03:54.3 & -43:20:56 & SBab  & SB     &    896& 10.4&  8.9$\times$5.6 &  $-$19.9& 271. &       0.24&    0.29&    9.77 &    \nodata &  \nodata &  53 -7      \cr
NGC~1566&  04:20:00.4 & -54:56:16 & SABbc & Sy     &   1496& 18.0&  8.3$\times$6.6 &  $-$21.9& 233. &       0.49&    0.32&    10.03 &   \nodata &  \nodata &  53 -1      \cr
NGC~1705&  04:54:13.5 & -53:21:40 & Am    & SB     &    628&  5.8&  1.9$\times$1.4 &  $-$16.7& \nodata &    0.29&    0.48&    \nodata & \nodata &  0.03    &  53 +1      \cr
NGC~2403&  07:36:51.4 & +65:36:09 & SABcd & H      &    131&  3.5& 21.9$\times$12.3&  $-$19.7& 257. &       0.29&    0.35&    9.73 &    7.86 &     1.3     &  14 -10     \cr
Ho~II   &  08:19:05.0 & +70:43:12 & Im    & \nodata&    157&  3.5&  7.9$\times$6.3 &  $-$17.1& 73. &        0.06&    0.44&    \nodata & \nodata &  0.12    &  14 -10     \cr
M81~DwA &  08:23:56.0 & +71:01:45 & I?    & \nodata&    113&  3.5&  1.3$\times$0.7 &  \nodata& 38. &     \nodata& \nodata&    $<$7.06 & 6.33 &     $<$0.001&  14 -10     \cr
DDO~053 &  08:34:07.2 & +66:10:54 & Im    & \nodata&     19&  3.5&  1.5$\times$1.3 &  $-$13.6& \nodata &    0.48&    0.29&    \nodata & \nodata &  0.005   &  14 -10     \cr
NGC~2798&  09:17:22.9 & +41:59:59 & SBa   & SB     &   1726& 24.7&  2.6$\times$1.0 &  $-$19.6& 316. &       4.71&    0.70&    9.29 &    9.47 &     2.0     &  21 -16     \cr
NGC~2841&  09:22:02.6 & +50:58:35 & SAb   & L/Sy   &    638&  9.8&  8.1$\times$3.5 &  $-$20.7& 611. &       0.11&    0.18&    9.53 &    9.30 &     0.2     &  15 +10     \cr 
NGC~2915&  09:26:11.5 & -76:37:35 & I0    & SB     &    468&  2.7&  1.9$\times$1.0 &  $-$15.1& 157. &       0.31&    1.17&    8.25 &    \nodata &  0.04    &  14 +20     \cr
Ho~I    &  09:40:32.3 & +71:10:56 & IABm  & \nodata&    143&  3.5&  3.6$\times$3.0 &  $-$13.2& 44. &     \nodata& \nodata&    8.15 &    \nodata &  0.004   &  14 -10     \cr
NGC~2976&  09:47:15.4 & +67:54:59 & SAc   & H      &      3&  3.5&  5.9$\times$2.7 &  $-$17.6& \nodata &    0.49&    0.37&    8.26 &    7.91 &     0.2     &  14 -10     \cr
NGC~3049&  09:54:49.6 & +09:16:18 & SBab  & SB     &   1494& 19.6&  2.2$\times$1.4 &  $-$18.7& 213. &       0.64&    0.67&    9.10 &    \nodata &  \nodata &  \nodata    \cr 
NGC~3031&  09:55:33.2 & +69:03:55 & SAab  & L      &  $-$34&  3.5& 26.9$\times$14.1&  $-$21.2& 446. &       0.08&    0.26&    8.88 &    \nodata &  1.1     &  14 -10     \cr 
NGC~3034&  09:55:52.2 & +69:40:47 & IO    & SB     &    203&  3.5& 11.2$\times$4.3 &  $-$17.9& \nodata &   42.35&    0.97&    8.85 &    9.39 &     6.0:    &  14 -10     \cr
Ho~IX   &  09:57:32.0 & +69:02:45 & Im    & \nodata&     46&  3.5&  2.5$\times$2.0 &  $-$13.6& \nodata & \nodata& \nodata&    7.55 &    \nodata &  0.001   &  \nodata    \cr
M81~DwB &  10:05:30.6 & +70:21:52 & Im    & \nodata&    350&  3.5&  0.9$\times$0.6 &  $-$12.5& 69. &     \nodata& \nodata&    7.08 &    \nodata &  0.004   &  14 -10     \cr
NGC~3190&  10:18:05.6 & +21:49:55 & SAap  & L      &   1271& 17.4&  4.4$\times$1.5 &  $-$20.7& 601. &       0.30&    0.32&    8.65 &    \nodata
 &  $<$0.1  &  21 -6      \cr
NGC~3184&  10:18:17.0 & +41:25:28 & SABcd & H      &    592&  8.6&  7.4$\times$6.9 &  $-$19.0& 142. &       0.99&    0.31&    9.32 &    8.96
&     1.2     &  15 +7      \cr
NGC~3198&  10:19:54.9 & +45:32:59 & SBc   & \nodata&    663&  9.8&  8.5$\times$3.3 &  $-$20.2& 318. &       0.19&    0.40&    9.74 &    \nodata
 &  0.85    &  15 +7      \cr
IC~2574 &  10:28:21.2 & +68:24:43 & SABm  & \nodata&     57&  3.5& 13.2$\times$5.4 &  $-$17.7& 123. &       0.11&    0.23&    9.21 &    \nodata
 &  0.10    &  14 -10     \cr
NGC~3265&  10:31:06.8 & +28:47:47 & E     & \nodata&   1421& 20.0&  1.3$\times$1.0 &  $-$17.7& \nodata &    1.30&    0.65&    8.24 &    \nodata &  \nodata &  \nodata    \cr
Mrk~33  &  10:32:31.9 & +54:24:03 & Im    & SB     &   1461& 21.7&  1.0$\times$0.9 &  $-$18.4& 181. &       2.29&    0.88&    8.77 &    \nodata &  1.5     &  13 +1      \cr 
NGC~3351&  10:43:57.7 & +11:42:13 & SBb   & SB     &    778&  9.3&  7.4$\times$5.0 &  $-$20.4& 288. &       0.54&    0.51&    9.10 &    8.82 &     1.2     &  15 -1      \cr
NGC~3521&  11:05:48.6 & -00:02:09 & SABbc & L      &    805&  9.0& 11.0$\times$5.1 &  $-$21.0& 466. &       0.81&    0.38&    9.67 &    9.64 &     1.7     &  15 -0 +1   \cr
NGC~3621&  11:18:16.3 & -32:48:45 & Sad   & \nodata&    727&  6.2& 12.3$\times$7.1 &  $-$19.4& 290. &       1.10&    0.33&    9.88 &    \nodata &  5.1     &  15 -0      \cr
NGC~3627&  11:20:15.0 & +12:59:30 & SABb  & Sy2    &    727&  8.9&  9.1$\times$4.2 &  $-$20.8& 378. &       1.21&    0.47&    8.92 &    9.61 &     6.9     &  15 -2 +1   \cr
NGC~3773&  11:38:13.0 & +12:06:43 & SA0   & \nodata&    987& 12.9&  1.2$\times$1.0 &  $-$17.5& 191. &       0.40&    0.82&    7.99 &    \nodata &  \nodata &  21 -4 +1   \cr
NGC~3938&  11:52:49.4 & +44:07:15 & SAc   & \nodata&    809& 12.2&  5.4$\times$4.9 &  $-$20.1& 112. &       0.46&    0.33&    9.57 &    9.46 &     1.2     &  12 -1      \cr
NGC~4125&  12:08:06.0 & +65:10:27 & E6p   & \nodata&   1356& 21.4&  5.8$\times$3.2 &  $-$21.6& \nodata &    0.03&    0.39&    \nodata & \nodata &  \nodata &  12 +5 -1   \cr
NGC~4236&  12:16:42.1 & +69:27:45 & SBdm  & \nodata&      0&  3.5& 21.9$\times$7.2 &  $-$18.1& 176. &       0.09&    0.40&    9.23 &    $<$8.14 &  0.3     &  14 +10     \cr
NGC~4254&  12:18:49.6 & +14:24:59 & SAc   & \nodata&   2407& 20.0&  5.4$\times$4.7 &  $-$21.6& 272. &       1.02&    0.37&    9.86 &    10.12 &    11.0    &  11 -1      \cr
NGC~4321&  12:22:54.9 & +15:49:21 & SABbc & L      &   1571& 20.0&  7.4$\times$6.3 &  $-$22.1& 283. &       0.73&    0.37&    9.67 &    10.17 &    5.5     &  11 -1      \cr
\enddata
\end{deluxetable}

\newpage

\begin{deluxetable}{lrrllrrrrrrrrrrl}
\rotate
\tabletypesize{\scriptsize}
\tablenum{2 (cont.)}
\tablecaption{SINGS Galaxy Sample}
\tablewidth{0pc}
\tablecolumns{16}
\tablehead{
\colhead{Galaxy} &
\colhead{RA} &
\colhead{Dec} &
\colhead{Type} & 
\colhead{Nuc} & 
\colhead{V$_r$} &
\colhead{Dist} &
\colhead{$D_{25}$} & 
\colhead{M$_{opt}$} & 
\colhead{W$_{20}$} &
\colhead{${{\rm L}_{IR}}\over{{\rm L}_{opt}}$} & 
\colhead{${{\rm F}_{60}}\over{{\rm F}_{100}}$} &
\colhead{$\log\,M_{H{\small I}}$} &
\colhead{$\log\,M_{H_2}$} &
\colhead{SFR} &
\colhead{Group} \cr
\colhead{} & 
\colhead{J2000} & 
\colhead{J2000} & 
\colhead{} &  
\colhead{} & 
\colhead{km/s} &
\colhead{Mpc} &
\colhead{arcmin} & 
\colhead{mag} & 
\colhead{km/s} &
\colhead{} &
\colhead{} & 
\colhead{M$_\odot$} &
\colhead{M$_\odot$} &
\colhead{M$_\odot$/yr} \cr
\colhead{} \cr 
\colhead{(1)}  &  \colhead{(2)}  &  \colhead{(3)}  &  \colhead{(4)}  &  \colhead{(5)}  &
\colhead{(6)}  &  \colhead{(7)}  &  \colhead{(8)}  &  \colhead{(9)}  &  \colhead{(10)} &
\colhead{(11)} &  \colhead{(12)} &  \colhead{(13)} &  \colhead{(14)} &  \colhead{(15)} &  \colhead{(16)} \cr 
}
\startdata
NGC~4450&  12:28:29.6 & +17:05:06 & SAab  & L      &   1954& 20.0&  5.2$\times$3.9 &  $-$21.4& 290. &       0.07&    0.19&    8.61 &    9.30 &     0.5     &  11 -1      \cr
NGC~4536&  12:34:27.1 & +02:11:16 & SABbc & H      &   1808& 25.0&  7.6$\times$3.2 &  $-$20.8& 337. &       2.33&    0.64&    9.71 &    9.71 &     3.7     &  11 -4 +1   \cr
NGC~4552&  12:35:39.9 & +12:33:22 & E0    & L      &    340& 20.0&  5.1$\times$4.7 &  $-$20.8& \nodata & \nodata& \nodata&    \nodata & \nodata &  \nodata &  11 -1      \cr
NGC~4559&  12:35:57.7 & +27:57:35 & SABcd & H      &    816& 11.6& 10.7$\times$4.4 &  $-$21.0& 251. &       0.17&    0.41&    10.05 &   \nodata &  \nodata &  14 -1      \cr
NGC~4569&  12:36:49.8 & +13:09:46 & SABab & L/Sy   & $-$235& 20.0&  9.5$\times$4.4 &  $-$22.0& 360. &       0.22&    0.37&    8.80 &    9.82 &     1.9     &  11 -1      \cr
NGC~4579&  12:37:43.6 & +11:49:05 & SABb  & L/Sy   &   1519& 20.0&  5.9$\times$4.7 &  $-$21.8& 390. &       0.17&    0.28&    8.91 &    9.60 &     2.0     &  11 -1      \cr
NGC~4594&  12:39:59.4 & -11:37:23 & SAa   & L/Sy2  &   1091& 13.7&  8.7$\times$3.5 &  $-$21.5& 762. &       0.16&    0.19&    8.77 &    $<$9.36 &  0.1     &  11 -14 +10 \cr
NGC~4625&  12:41:52.7 & +41:16:25 & SABmp & \nodata&    609&  9.5&  2.2$\times$1.9 &  $-$17.5& 86. &        0.37&    0.34&    9.02 &    \nodata &  \nodata &  14 -4      \cr
NGC~4631&  12:42:08.0 & +32:32:26 & SBd   & \nodata&    606&  9.0& 15.5$\times$2.7 &  $-$20.6& 320. &       1.26&    0.40&    10.09 &   9.19 &     3.3     &  14 -6      \cr
NGC~4725&  12:50:26.6 & +25:30:03 & SABab & Sy2    &   1206& 17.1& 10.7$\times$7.6 &  $-$22.0& 410. &       0.09&    0.20&    9.87 &    9.80 &     \nodata &  14 -2 +1   \cr
NGC~4736&  12:50:53.0 & +41:07:14 & SAab  & L      &    308&  5.3& 11.2$\times$9.1 &  $-$19.9& 241. &       0.87&    0.57&    8.94 &    8.90 &     2.1     &  14 -7      \cr 
DDO~154 &  12:54:05.2 & +27:08:59 & IBm   & \nodata&    376&  5.4&  3.0$\times$2.2 &  $-$15.1& 103. &    \nodata& \nodata&    8.99 &    \nodata &  0.0015  &  14 +3      \cr
NGC~4826&  12:56:43.7 & +21:40:52 & SAab  & Sy2    &    408&  5.6& 10.0$\times$5.4 &  $-$20.3& 311. &       0.27&    0.48&    8.49 &    8.87 &     0.3     &  14 +3      \cr
DDO~165 &  13:06:24.8 & +67:42:25 & Im    & \nodata&     37&  3.5&  3.5$\times$1.9 &  $-$15.3& 68. &     \nodata& \nodata&    8.02 &    \nodata &  0.002   &  14 +10     \cr
NGC~5033&  13:13:27.5 & +36:35:38 & SAc   & Sy2    &    875& 13.3& 10.7$\times$5.0 &  $-$20.9& 446. &       0.48&    0.34&    9.97 &    9.50
&     2.1     &  43 -1      \cr
NGC~5055&  13:15:49.3 & +42:01:45 & SAbc  & H/L    &    504&  8.2& 12.6$\times$7.2 &  $-$19.0& 405. &       4.56&    0.25&    9.88 &    9.62
&     2.3     &  14 -5      \cr
NGC~5194&  13:29:52.7 & +47:11:43 & SABbc & H/Sy2  &    463&  8.2& 11.2$\times$6.9 &  $-$21.4& 195. &       0.60&    0.35&    9.60 &    9.83
&     5.4     &  14 -5      \cr
NGC~5195&  13:29:58.7 & +47:16:05 & SB0p  & L      &    552&  8.2&  5.8$\times$4.6 &  $-$20.0& \nodata &    0.29&    0.53&    \nodata & 8.25
&     $<$0.1  &  14 -5      \cr
NGC~5398&  14:01:21.3 & -33:03:47 & SBdm  & H      &   1216& 15.0&  2.8$\times$1.7 &  $-$18.9& 137. &       0.44&    0.58&    9.11 &    \nodata
 &  \nodata &  16 -5      \cr
NGC~5408&  14:03:20.9 & -41:22:40 & IBm   & \nodata&    509&  4.5&  1.6$\times$0.8 &  $-$16.1& 114. &       0.74&    0.89&    8.45 &    \nodata
 &  0.18    &  14 -15     \cr
NGC~5474&  14:05:01.6 & +53:39:44 & SAcd  & H      &    273&  6.9&  4.8$\times$4.3 &  $-$18.4& 61. &        0.11&    0.28&    9.10 &    \nodata
 &  0.2     &  14 -9      \cr
NGC~5713&  14:40:11.5 & -00:17:21 & SABbcp& \nodata&   1883& 26.6&  2.8$\times$2.5 &  $-$20.9& 209. &       1.71&    0.57&    9.93 &    9.72
&     \nodata &  41 -2 +1   \cr
NGC~5866&  15:06:29.5 & +55:45:48 & S0    & \nodata&    692& 12.5&  4.7$\times$1.9 &  $-$19.9& \nodata & 0.51&    0.29&    $<$8.28 & 8.63 &     $<$0.1  &  44 -1      \cr
IC~4710 &  18:28:38.0 & -66:58:56 & SBm   & SB     &    741&  8.5&  3.6$\times$2.8 &  $-$18.3& 51. &  0.20&    0.38&    8.69 &    \nodata &  1.3     &  19 -1      \cr
NGC~6822&  19:44:56.6 & -14:47:21 & IBm   & \nodata&  $-$57&  0.6& 15.5$\times$13.5&  $-$13.8& 81. &  2.50&    0.50&    8.26 &    \nodata &  0.024   &  14 -12     \cr
NGC~6946&  20:34:52.3 & +60:09:14 & SABcd & H      &     48&  5.5& 11.5$\times$9.8 &  $-$21.3& 242. &  0.39&    0.40&    9.79 &    9.61 &     2.2     &  14 -0      \cr
NGC~7331&  22:37:04.1 & +34:24:56 & SAb   & L      &    816& 15.7& 10.5$\times$3.7 &  $-$21.8& 530. &  1.02&    0.30&    10.01 &   10.05 &    4.2     &  65 -1      \cr
NGC~7552&  23:16:11.0 & -42:34:59 & SAc   & SB/L   &   1585& 22.3&  3.4$\times$2.7 &  $-$21.7& 280. &   3.20&    0.71&    9.68 &    \nodata &  7.0     &  61 -16     \cr
NGC~7793&  23:57:49.8 & -32:35:28 & SAd   & H      &    230&  3.2&  9.3$\times$6.3 &  $-$18.2& 194. &   0.57&    0.35&    8.81 &    \nodata &  0.3     &  14 -13     \cr
\enddata
\tablecomments{Col (1): ID; 
Col (2): The right ascension in the J2000.0 epoch;
Col (3): The declination in the J2000.0 epoch;
Col (4): RC3 Type;
Col (5): Nuclear Type: H=HII-region, SB=Starburst, L=LINER, Sy=Seyfert(1, 2);
Col (6): Heliocentric velocity;
Col (7): Flow-corrected distance in Mpc, for H$_0$ = 70 km\,s$^{-1}$\,Mpc$^{-1}$
Col (8): Major and minor axis diameters;
Col (9): Absolute $R$ magnitude, when available; otherwise from the $V$ or $B$ bands;
Col (10): 21\,cm neutral hydrogen line width at 20\% of maximum intensity, in
km s$^{-1}$, as given in Tully (1988) or RC3;
Col (11): FIR/optical luminosity ratio.  The FIR luminosity is derived from 
the IRAS-measured 60--100 \um\ fluxes, from Fullmer \& Lonsdale (1989), 
the optical luminosity is defined as: $L_{opt} \propto {D^2} f_\nu$;
Col (12): The ratio of the IRAS 60\,\um\ to 100\,um\ flux;
Col (13): The logarithmic atomic gas mass, from H\,I integrated fluxes;
Col (14): The logarithmic molecular gas mass, from CO integrated fluxes;
Col (15): Star formation rates derived from \Ha\ emission, with  
typical $A(H\alpha)$  = 1 mag;
Col (16): Group affiliation code from Tully (1988).  Important codes
include 11 -1 (the Virgo Cluster) and 14 -10 (the M~81 Group).}
\end{deluxetable}

\newpage

\begin{deluxetable}{lrrcrl}
\tabletypesize{\footnotesize}
\tablecaption{Summary of SIRTF Imaging} 
\tablewidth{0pc}
\tablenum{4a}
\tablecolumns{3}
\tablehead{
\colhead{Instrument} & \colhead{Wavelength} & \colhead{Bandwidth} &
\colhead{Aereal Coverage} & \colhead{Resolution} & \colhead{Sensitivity\tablenotemark{a}} \cr 
\colhead{} & \colhead{$\mu$m} & \colhead{$\mu$m} & \colhead{arcmin} & 
\colhead{arcsec} & \colhead{MJy/sr} \cr
}
\startdata
IRAC & 3.6 & 0.75 &  $>$D$_{25}$\tablenotemark{b} & 1.7 & ~~~0.02 \cr
     & 4.5 & 1.02 &  $>$D$_{25}$\tablenotemark{b} & 1.7 & ~~~0.03 \cr
     & 5.8 & 1.44 &  $>$D$_{25}$\tablenotemark{b} & 1.7 & ~~~0.09 \cr
     & 8.0 & 2.91 &  $>$D$_{25}$\tablenotemark{b} & 2.0 & ~~~0.12 \cr
\\
\tableline
\\
MIPS & 24 & 4.7 &  $>$D$_{25}$\tablenotemark{b} & 5.7 & ~~~0.2 \cr
     & 70 & 19. &  $>$D$_{25}$\tablenotemark{b} & 17. & ~~~0.5 \cr
     & 160 & 35. &  $>$D$_{25}$\tablenotemark{b} & 38. & ~~~0.5 \cr
\enddata
\tablenotetext{a}{3-$\sigma$ surface brightness limit for average background
target, based on pre-flight instrument performance estimates.  
Variations in local background can alter these values up to factor of 2.}
\tablenotetext{b}{Full coverage of galaxy to at least RC3 D$_{25}$ 
or 5\arcmin\ $\times$ 5\arcmin, whichever is greater}
\end{deluxetable}

\begin{deluxetable}{lcccc}
\tabletypesize{\footnotesize}
\tablecaption{Summary of SIRTF Spectroscopy}
\tablewidth{0pc}
\tablenum{4b}
\tablecolumns{3}
\tablehead{
\colhead{Instrument/Mode} & \colhead{Wavelengths} & \colhead{Resolution} &
\colhead{Aereal Coverage} & \colhead{Sensitivity\tablenotemark{a}} \cr
\colhead{} & \colhead{$\mu$m} & \colhead{$R$} & \colhead{arcmin} & 
\colhead{3-$\sigma$} \cr
}
\startdata
MIPS SED & 55 $-$ 96 & 15 & 1.0 $\times$ 0.55 D$_{25}$\tablenotemark{b} & 1.2 MJy/sr\tablenotemark{e} \cr
IRS SED  & 14 $-$ 40 & 62 $-$ 124 & 0.9 $\times$ 0.55 D$_{25}$\tablenotemark{b} & 0.6 MJy/sr\tablenotemark{f} \cr
\\
\tableline
\\
IRS Nuclear & 5.3 $-$ 14.2 & 62 $-$ 124 & 0.3 $\times$ 0.9 & 3 mJy\tablenotemark{g} \cr
	    & 10.0 $-$ 37.0 & 600 & 0.3 $\times$ 0.4\tablenotemark{d} & 
	    5\,$\times$\,10$^{-18}$ W/m$^2$\tablenotemark{h} \cr
\\
\tableline
\\
IRS Extra-nuclear & 5.3 $-$ 14.2 & 62 $-$ 124 & 0.3 $\times$ 0.9 & 3 mJy\tablenotemark{g} \cr
	   & 10.0 $-$ 37.0 & 600 & 0.3 $\times$ 0.4\tablenotemark{d} &
	   5\,$\times$\,10$^{-18}$ W/m$^2$\tablenotemark{h}   \cr
\enddata
\tablenotetext{a}{3-$\sigma$ sensitivities based on pre-flight estimates,
see other footnotes for remarks on individual observing modes.}
\tablenotetext{b}{Approximate radial coverage, see \S 4.3 for details.  
Minimum strip length is 4\arcmin.}
\tablenotetext{c}{Approximate radial coverage, see \S~4.4.1 for details.
Minimum strip length is 2\farcm5.} 
\tablenotetext{d}{Spatial coverage given for 10.0 $-$ 19.5~$\mu$m channel.
Maps at 19.3 $-$ 37.0~$\mu$m are 0\farcm55 $\times$ 0\farcm75, to account
for degraded spatial resolution at the longer wavelengths.}
\tablenotetext{e}{Continuum surface brightness sensitivity}
\tablenotetext{f}{Continuum surface brightness sensitivity at 15\,\micron}
\tablenotetext{g}{Point source continuum flux sensitivity}
\tablenotetext{h}{Point source sensitivity limit for unresolved emission line}
\end{deluxetable}

\begin{deluxetable}{lll}
\tablecaption{SINGS SIRTF Data Products}
\tablewidth{0pt}
\tablenum{6}
\tablecolumns{3}
\tablehead{
\colhead{Instrument} &
\colhead{Rapid Release Products} & \colhead{Enhanced Release Products} \cr
}
\startdata
IRAC & Mosaic Images & Full-Resolution Dithered Images \cr
MIPS Image & Full-Map Images  & MIPS Enhancer Images \cr
MIPS SED &  1D Spectra &  MIPS Enhancer Spectral Cubes \cr
IRS  & 1D Spectra, Line Maps & 3D Spectral Cubes \cr
\enddata
\end{deluxetable}

\begin{deluxetable}{lcccrcl}
\rotate
\tablecaption{Summary of SINGS Ancillary and Complementary Data} 
\tabletypesize{\scriptsize}
\tablewidth{0pc}
\tablenum{5}
\tablecolumns{3}
\tablehead{
\colhead{Type} &
\colhead{Wavelength} & \colhead{Coverage} & \colhead{Resolution} & 
\colhead{Galaxies\tablenotemark{a}} & \colhead{Type\tablenotemark{b}} & 
\colhead{Facilities\tablenotemark{c}} \\
\colhead{} & \colhead{} & \colhead{arcmin} & \colhead{arcsec} & \colhead{} & \colhead{} & \colhead{} \\
\colhead{(1)} & \colhead{(2)} & \colhead{(3)} & \colhead{(4)} & \colhead{(5)} 
& \colhead{(6)} & \colhead{(7)} \\   
}
\startdata
High-Res HI Maps & 21\,cm & 30 & 7 & 35 & Comp & VLA \\
Low-Res HI Maps &  21\,cm & 30 & 30 $-$ 45 & $\sim$70 & Arch, SINGS & VLA, WHISP, ATCA \\
HI Maps (WSRT) & 21\,cm & 30 & 12 $\times$ 12/$\sin \delta$ & 30 & Comp & WSRT \\
Radio Continuum Maps (WSRT) & 18\,cm, 22\,cm & 30 &  12 $\times$ /$\sin \delta$ & 30 & Comp & WSRT \\
Radio Continuum Maps (VLA) & 20\,cm & 30 & 30 $-$ 45 & $\sim$60 & SINGS & VLA \\
Radio Continuum Maps (ATCA) & 20\,cm & \nodata & 10 $\times$ 10/$\sin \delta$ & 7 & SINGS & ATCA \\
High-Res CO Maps & 1.3\,mm & 1.7 $-$ 3.2 & 7 & 20 & Arch & BIMA SONG \\
Low-Res CO Maps & 1.3\,mm & 3 $-$ 5 & 50 & $\sim$40 & Comp & FCRAO \\
Submillimeter Maps & 850, 450\,\micron & 2.3 $-$ 10 & 7.5 $-$ 14 & $\ge$20 & Arch, Comp & JCMT (SCUBA) \\
FIR Photometry & 170 $-$ 240\,\micron\ & \nodata & 90 & 30 & Arch & ISO (PHT) \\
$JHK_s$ Imaging (2MASS) & 1.2, 1.6, 2.2\,\micron\ & $\ge D_{25}$ & 2 $-$ 3 & 75 & Arch & 2MASS \\
$JHK_s$ Imaging (Deep) & 1.2, 1.6, 2.2\,\micron\ & 2.5 $-$ 8.5 & 1 $-$ 2 & 75 & SINGS & Steward, CTIO, Palomar \\
Paschen\,$\alpha$, $H$ Imaging & 1.88, 1.6\,\micron\ & 0.8 & 0.2 & $\sim$50 & SINGS & HST (NICMOS) \\ 
H$\alpha$ Imaging & 0.656\,\micron\ & $\ge D_{25}$ & 1 $-$ 3 & 75 & SINGS & KPNO, CTIO, Steward \\
H$\alpha$ Imaging/Kinematics & 0.656\,\micron\ & 4 $-$ 20 & 1 $-$ 3 & $\ge$30 & Comp & CFHT, ESO, OHP, OMM \\
$BVRI$ Imaging & 0.44, 0.55, 0.64, 0.81\,\micron\ & $\ge D_{25}$ & 1 $-$ 2 & 75 & SINGS & KPNO, CTIO \\
UV Imaging & 0.15, 0.25\,\micron\ & $\ge D_{25}$ & 3 & 27 & Arch & UIT \\
UV Imaging & 0.16, 0.24\,\micron\ & $\ge D_{25}$ & 5 & 75 & Comp & GALEX \\
X-Ray Imaging & 0.5 $-$ 8 keV & 17 & 2 & $\ge$30 & Arch & CXO \\
\tableline
Spectrophotometry (strips) & 0.36 - 0.69\,\micron\ & 0.9 $\times$ 0.5 $D_{25}$ & \nodata & 75 & SINGS & Steward, CTIO \\
Spectrophotometry (centers)  & 0.36 - 0.69\,\micron\ & 0.33 $\times$ 0.33 & \nodata & 75 & SINGS &  Steward, CTIO \\
Spectrophotometry (nuclei)  & 0.36 - 0.69\,\micron\ & 0.04 $\times$ 0.04 & \nodata &  75 & SINGS & Steward, CTIO \\
UV Spectra (centers) & 0.12 $-$ 0.32\,\micron\ & 0.17 $-$ 0.33 & \nodata & 25 & Arch & IUE \\
\enddata
\tablenotetext{a}{Approximate number of galaxies for which data will be 
available when projects are completed.}  
\tablenotetext{b}{Type of dataset: Arch = archival data (in some cases reprocessed by the SINGS team; SINGS = ancillary data being obtained by members of the SINGS team; Comp = complementary data being obtained by independent groups, but with plans to archive when projects are completed.}
\tablenotetext{c}{Facilities as follows:  ACTA = Australia Telescope Compact
Array; BIMA = Berkeley-Illinois-Maryland Array (BIMA SONG survey); 
CFHT = Canada France Hawaii 3.6\,m Telescope; 
CTIO = Cerro Tololo Interamerican Observatory 1.5\,m telescope; 
CXO = Chandra X-Ray Observatory; ESO = European Southern Observatory 3.6\,m
telescope; 
FCRAO = Five College Radio Astronomy Observatory; GALEX = Galaxy Evolution
Explorer; HST = Hubble Space Telescope (NICMOS instrument); 
ISO: Infrared Space Observatory, ISOPHOT instrument; 
IUE: International Ultraviolet Explorer; JCMT = James Clerk Maxwell Telescope;
KPNO = Kitt Peak National 
Observatory 2.1\,m telescope; OHP = Observatoire de Haute Provence 1.93\,m 
telescope; OMM = Observatoire Mont Megantic 1.6\,m telescope; 
Palomar: Palomar Observatory 5.0\,m telescope; 
Steward = Steward Observatory Bok 2.3\,m and
Bigelow 1.5\,m telescopes; UIT = Ultraviolet Imaging Telescope; 
VLA = NRAO Very Large Array; WHISP = Westerbork Observations of Neutral
Hydrogen in Irregular and Spiral Galaxies Survey; 
WSRT = Westerbork Synthesis Radiotelescope} 
\end{deluxetable}
			      
\end{document}